\documentclass[journal]{vgtc}              %
\usepackage{booktabs}
\usepackage{float}
\usepackage{longtable}
\usepackage{xcolor}
\usepackage[table]{xcolor}
\usepackage{wrapfig}
\newcommand{\rev}[1]{\textcolor{black}{#1}}
\usepackage{graphicx}
\usepackage{array}
\usepackage{adjustbox}
\newcolumntype{I}{!{\vrule width 1.5pt}}
\usepackage{tikz}
\usetikzlibrary{calc}
\usepackage{tabularx}
\usepackage{ragged2e}
\newcolumntype{N}[1]{>{\hyphenpenalty=10000\exhyphenpenalty=10000\raggedright\arraybackslash}p{#1}}

\onlineid{1733}
\usepackage{amsmath}

\vgtccategory{Research}

\title{Mixed Uncertainty in One View:\\Co-Visualizing Statistical Variability and Qualitative Confidence}
\author{%
  \authororcid{Racquel Fygenson}{0000-0002-0705-9000},
  \authororcid{Lace Padilla}{0000-0001-9251-5279}, and 
  \authororcid{Laura E. Matzen}{0000-0002-7720-2896}
}

\authorfooter{
  \item
  	Racquel Fygenson and Lace Padilla are with Northeastern University.
  	E-mail: fygenson.r | l.padilla @northeastern.edu
  \item
  	Laura E. Matzen is with Sandia National Laboratories
  	E-mail: lematze@sandia.gov.

}

\abstract{Forecasting involves multiple forms of uncertainty, including both uncertainties that can be quantified directly (quantitative uncertainty) and those that must be expressed through experts’ subjective judgments about the forecast and its context (qualitative confidence).
 Past work has established that conveying both quantitative uncertainty and qualitative confidence in forecasts can \rev{alter} readers\rev{' decision making}, but little research investigates the impact of how these forms of uncertainty are presented. In this work, we present three preregistered human-subjects studies (total \textit{n} = 923) on how different methods of visualizing qualitative uncertainty alongside line charts' confidence intervals affects \rev{non-experts}' decision making. In particular, we investigate representing qualitative uncertainty separately via text and icons, and integrated into quantitative confidence intervals via color, transparency, and a blurred stroke design. In Experiment 1, we confirm that showing qualitative confidence alongside statistical variability can change patterns of decision making, replicating findings from previous work in the new context of time-series line charts. In Experiments 2 and 3, we find several non-textual encoding techniques that produce similar effects in participants' incorporation of  qualitative confidence into their judgments. Our findings suggest actionable guidelines for visualization designers who seek to represent multiple forms of uncertainty for a single line chart forecast. A free copy of this paper and all supplemental materials are available at https://osf.io/7ya2c/overview.
 }


\teaser{
  \centering
  \includegraphics[width=\linewidth, alt={An overview of the experimental stimuli showing an example decision task alongside the qualitative confidence encodings used in the study, including text, glyph-based, and overlaid conditions implemented with transparency, color, and blur.}]{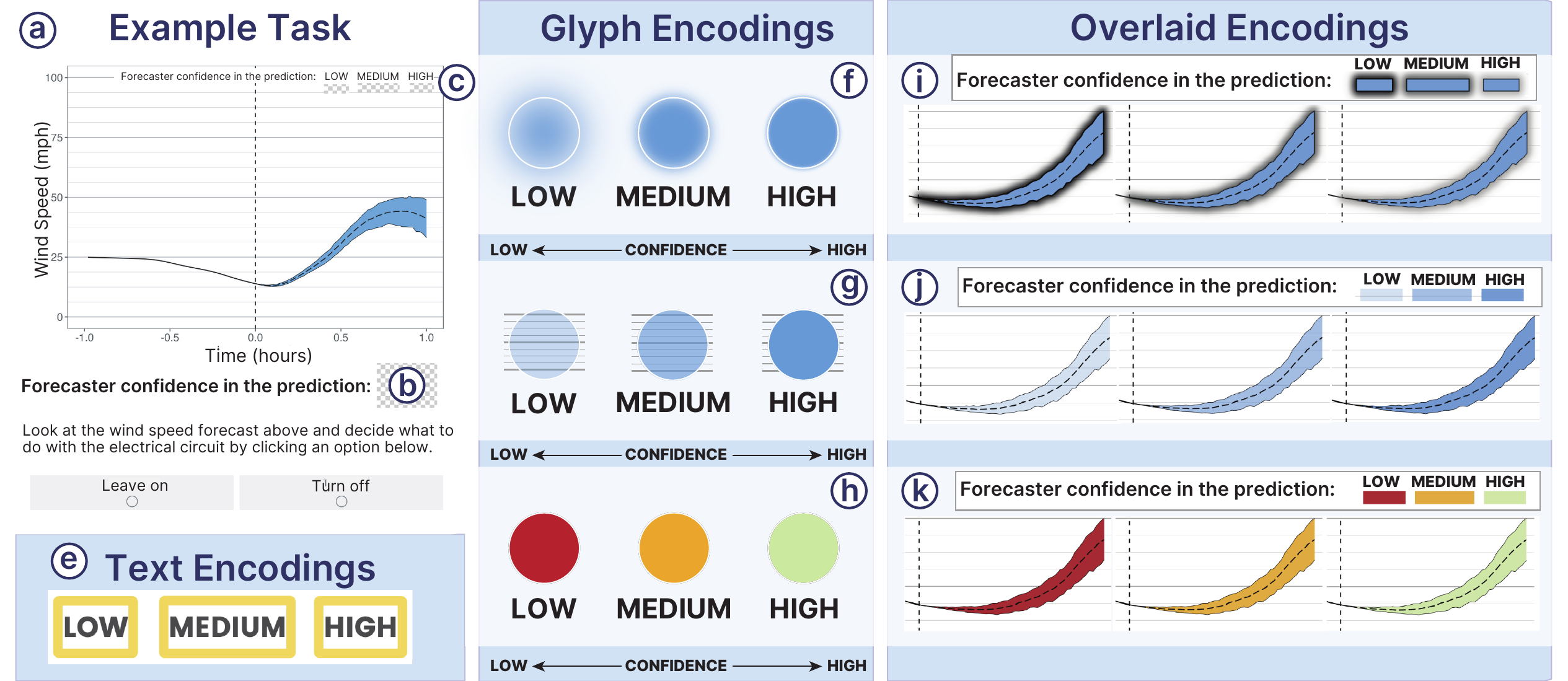}
  \caption{%
    \rev{(a) an example experimental trial asking participants to decide to leave on a circuit or turn it off based on estimated wind speed forecast. (b) checkered box shows location of low, medium, or high text (e) or glyph encoding (f, g, h) which is used to communicate forecaster confidence as label under the forecast in Experiments 1 and 2. (c) checkered boxes shows where legends are placed, when overlaid encodings (i, j, k) are used to communicate forecaster confidence by integrating it directly on top of forecasts in Experiment 3. (f) and (i) show blur, (g) and (j) show transparency, and (h) and (k) show color hue conditions.} %
  }
  \label{fig:teaser}
}

\graphicspath{{figs/}{figures/}{pictures/}{images/}{./}} %

\usepackage{tabu}                      %
\usepackage{booktabs}                  %
\usepackage{lipsum}                    %
\usepackage{mwe}                       %
\usepackage{ccicons}                   %

\usepackage{mathptmx}                  %

\begin{document}

\firstsection{Introduction}

\maketitle
Uncertainty enters forecasting and modeling pipelines in many forms~\cite{pang1997approaches}, and numerous frameworks have been proposed to classify these uncertainties~\cite{van2019communicating}. These frameworks distinguish, for example, among \emph{aleatoric} uncertainty, which reflects irreducible randomness inherent in a process; \emph{epistemic} uncertainty, which arises from incomplete knowledge and, in principle, may be reduced; and \emph{ontological} uncertainty, which concerns how accurately a model represents reality~\cite{walker2003defining,derKiureghian2009aleatory,spiegelhalter2017risk}. In line with these distinctions, some uncertainty can be quantified directly (e.g., through statistical variability%
~\cite{van2019communicating}), whereas other uncertainty is harder to quantify formally and instead reflects expert judgments about evidence quality, model adequacy, or forecast reliability in context. In this paper, we use the terms \emph{quantitative} and \emph{qualitative} uncertainty to describe the practical distinction for visualization: \emph{quantitative uncertainty} can be expressed numerically, whereas \emph{qualitative uncertainty} cannot, and is instead often captured through expert judgment expressed in a non-numeric format. 

Within data visualization, a large body of work has identified successful methods for communicating quantitative uncertainty (for review, see \cite{Padilla2020review}). However, only a few studies have considered methods for communicating forecasts' qualitative uncertainty. In one study, researchers modified the shape of a forecast distribution to better align viewers’ subjective probability judgments with researchers' intended probabilities%
~\cite{yang2023subjective}. In a second paper, authors expressed experts' uncertainty about a model's reliability using textual labels such as ``low'', ``medium'', and ``high forecaster confidence''~\cite{padilla-2021-unc-about-unc}. This work found that members of the general public will generally use textual expressions of expert confidence to update their forecast-based decision making.%

\rev{Motivated by prior findings, we investigate how visually presenting qualitative confidence alongside quantitative uncertainty can impact decision making. Visualizing both pieces of information allows for more design freedom, and thus a variety of potential differences in decision making. For example, layering both pieces of information into a single visual object, an opportunity that is not afforded by textual communication, could impact the order, or mechanisms, with which readers integrate multiple forms of uncertainty into a single decision. Additionally, visual design can manifest in more diverse output options than text, which could also impact decision making.}
\rev{Thus, despite prior work establishing the existence of \textit{an} impact of qualitative confidence on decision making,} there remains little guidance for forecasters to express confidence along with, or as part of, \rev{their} %
visualization\rev{s}. 

Such guidance is especially relevant to applied settings, in which decision makers require quick access to multiple kinds of uncertainty to make important, time-sensitive decisions. For example, emergency management personnel in wildfire-prone areas need to monitor time-series weather forecasts and make quick decisions to turn off electrical energy grids before wind speed crosses a safety threshold and potentially ignites a fire ~\cite{vazquez-2022-wildfire-mitigation, cpuc-psps-website, huang-psps-review-2023}. This forecast-informed decision is crucial to millions of lives in at-risk regions, such as California~\cite{cpuc-psps-website}, and precludes deterministic solutions, instead requiring personnel to interpret multiple forms of uncertainty to make a decision~\cite{huang-psps-review-2023}.
At present, however, it is unclear which uncertainty presentation methods, if any, are effective in a time-series forecast context, and whether some designs may be preferable to others. 

To address this gap, we conduct three preregistered studies that examine a broad set of methods for communicating expert confidence alongside quantitative uncertainty. \rev{We do so through a controlled wind speed forecasting task, in which non-experts are faced with electricity shut-off decisions.} We compare text with two classes of visual confidence encodings: \textbf{glyph-based encodings} positioned below forecast charts and \textbf{overlaid encodings} applied directly to the forecasts. For both visual approaches, we test variances of blur, transparency, and color as candidate encodings for qualitative confidence, evaluating how alternative visual expressions of confidence shape decisions under uncertainty. 

The key contributions of this work include: 
\begin{itemize}
    \item Replication of prior findings that viewers spontaneously incorporate subjective forecaster confidence into decision-making~\cite{padilla-2021-unc-about-unc} and extension to a time-series forecast context.
    
    \item Evaluation of six previously untested visual encodings for communicating subjective forecaster confidence, showing that a subset performs comparably to textual confidence expressions.
    
    \item Empirical examination of the reported intuitiveness of tested visual encodings in communicating uncertainty.%
    
    \item Practical design guidance for communicating qualitative expert confidence alongside quantitative forecast uncertainty.
\end{itemize}

\section{Background}
\subsection{\rev{Visualizing} Qualitative Uncertainty}
Quantitative uncertainty is often expressed numerically, for example, through statistical variability, measurement error, or probabilistic ranges%
~\cite{Padilla2020review}. In contrast, qualitative uncertainty reflects forecast aspects that are difficult to quantify directly but that experts can still judge subjectively. For example, an expert may have greater confidence in a flood forecast for a region with stronger monitoring infrastructure%
, or less confidence in a model that omits a relevant factor, such as seasonality.

Although extensive research has focused on communicating quantitative forecast uncertainty (for reviews, see~\cite{heggli2023visual, Padilla2020review, matzen2023visualizing}), qualitative forecast uncertainty has received comparatively less attention.
In this space, Boukhelifa et al. explored sketchiness, dashing, grayscale, and blur for encoding qualitative uncertainty into the lines of different visualizations\cite{Boukhelifa2012-sketchiness}. Unfortunately, %
sketchiness, dashing, and blur can interfere with marks' exact edge placement and thus may alter confidence intervals' (CIs) positional encoding. %

Qualitative uncertainty can also be represented by ordinal uncertainty visualization if expressed using preset levels (e.g., "The forecast is [definitely inaccurate/probably accurate]"). MacEachren et al. investigated ordinal uncertainty encodings to determine their intuitiveness in communicating the concept of uncertainty~\cite{maceachren2012visual}. Their work suggests that blur and location (i.e., distance from the center of a target) are the two most intuitive forms of uncertainty representation, but only explores visual encodings in glyph form, not applied to charts' marks.%

 Lack of research of qualitative uncertainty representation is a notable limitation because there are important reasons for forecasters to communicate non-numeric uncertainty. %
 First, readers do not treat forecasts as perfectly certain when uncertainty information is omitted. Instead, they often infer additional uncertainty based on their own assumptions, which may vary widely and fail to align with available evidence~\cite{Morss-2008-unc, joslyn-2010-pubperceptionunc}. Second, qualitative information about forecasts, such as a their reliability in context~\cite{burgeno-2023-forecastquality}, may be relevant for important decision making. %
 Omitting qualitative uncertainty forces audiences to rely on incomplete forecast representations. Such an omission does not eliminate qualitative uncertainty's influence, it merely obscures its effects.
 
\subsection{Visualizing Multidimensional Uncertainty}
A large body of work has focused on communicating quantifiable, one-dimensional  uncertainty, yielding effective techniques such as quantile dotplots~\cite{kay2016ish}, violin plots~\cite{correll2014error}, and hypothetical outcome plots~\cite{hullman2015hypothetical}. By comparison, %
multidimensional uncertainty has received less attention. Prior work has focused on ensemble-based data, particularly in forecasting contexts such as weather and flooding %
(for review see, \cite{potter2025navigating}). 
One class of visualization approaches depicts some or all of the underlying ensemble members used to generate a forecast. Another class, display multiple conflicting forecasts, each from a different source, to communicate quantitative uncertainty implicitly~\cite{padilla2022multiple}. Both of these classes display multiple outcomes and can support judgments about spread and shape, but require viewers to \textit{infer} most likely outcomes~\cite{potter2025navigating, padilla2022multiple}. %

A third class summarizes ensembles into higher-level statistical displays. These visualizations communicate aggregate properties such as central tendency or spread, representing uncertainty more explicitly~\cite{potter2025navigating}, which can lead to incorrect inferences of determinism~\cite{joslyn2021visualizing}. Examples include median forecasts, CIs, and contour boxplots.%

Despite their limitations (e.g.,~\cite{joslyn2021visualizing, correll2014error}), summary techniques may remain attractive in applied settings because they are compact, familiar, and visually simple. For example, prior research has found that summary displays are more trusted than charts that show several forecasts at once~\cite{padilla2022multiple}. In our experiments, we use a summary-based representation to communicate quantitative forecast uncertainty. This choice is especially useful for our experimental goals because summary displays are less visually complex, and thus more easily support additional information encoding. We overlay subjective forecaster confidence onto a CI using color hue, blur, and transparency, to study how quantitative and qualitative uncertainty can be communicated in one visual.

\subsection{Decision Making in Vis Studies}
\label{sec:dec_vis}
Decision making is a common metric for evaluating the effects of visualization design~\cite{padilla2018decision}. Such evaluations can illuminate the ways people utilize visual data to operate. However, recent work cautions that decision studies can lead to inaccurate conclusions when they lack fully defined decision problems, such that participants are not provided with all the information necessary to make a rational decision \cite{Hullman-2025-underspecified}. Additionally, other literature advocates for the importance of differentiating between decision tasks and judgments when describing experimental outcomes and extrapolating findings~\cite{oral2023decoupling}. This literature characterizes decision tasks as forced-choice and future-oriented, accompanied by actions, and carrying a personal stake for the decision maker.

In this paper, we use a well-specified decision task with a clear incentive structure, explicit decision threshold, and stated interpretation of our stimuli to create a full defined decision problem. Our study participants also completed comprehension checks to ensure they were aware of this decision problem before beginning experimental trials. %

\subsection{Public Safety Power Shutoffs}
\label{sec:psps}
Public safety power shutoffs (PSPS), the context used in our experimental trials, are %
decisions that require synthesizing uncertainty and weighing large, tangible risks. During a PSPS, an electric utility shuts off power to parts of their grid to reduce wildfire risk in areas where strong winds could bring trees into power lines and ignite fires~\cite{huang-psps-review-2023}. Although PSPS events can lower fire danger, they disrupt customers and civic functions and are used as a last resort~\cite{huang-psps-review-2023}. Management personnel must make PSPS decisions using uncertain weather forecasts, in short time frames, and often without ideal information. %

Prior visualization research has used mock PSPS tasks to evaluate how uncertainty impacts visualization-based decision-making, showing that participants’ decisions were highly sensitive to the method with which quantitative uncertainty was represented~\cite{Matzen-2024-psps, matzen2025preprint}. For example, participants made more risk-averse decisions when forecast uncertainty was shown with a 95\% confidence interval (CI) than with a 50\% CI, suggesting that forecast readers can rely on visual heuristics, such as whether a CI crosses a decision-making threshold. In our experiments, we use a similar mock PSPS task to prior work~\cite{Matzen-2024-psps, matzen2025preprint}.

\begin{figure}[t]
  \centering
\includegraphics[width=\columnwidth, alt={A six-panel figure showing the six forecast distribution shapes used to create stimuli in the three experiments.}]{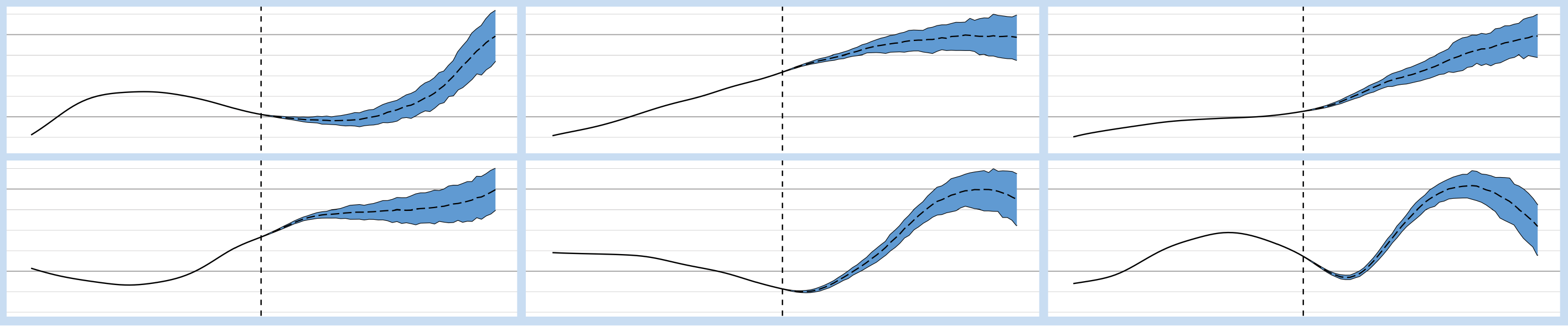}
  \caption{%
    The six forecast distribution shapes used in our experiments.%
  }
\label{fig:six_forecast_shapes}
  \vspace{-1em}
\end{figure}

\section{Materials and Methods}

\subsection{Experimental Task, Stimuli, and Reward Scheme}
\label{sec:taskstim}

\noindent\textbf{Task.} We present three preregistered experiments, all approved by Northeastern University's IRB (No. 23-07-18), and all with similar instructions and tasks. Participants were told that they would be ``playing the role of an electric grid operator who is monitoring an electrical circuit in an area where there is high fire danger,''  that we would be showing them graphs of past and future wind speed, that these graphs were uncertain, and that they should ``try to keep the circuit on when [they] can, and turn it off when [they] think there's too much risk of the wind speed going above 50 mph.'' Participants were shown example forecasts and informed about how they were generated, assumptions they could make about the forecasts, the source of potential uncertainty in these forecasts, and a clear monetary bonus incentive scheme. Throughout these instructions, participants completed comprehension checks to confirm they understood the instructions before they could proceed further. For each trial, the participants saw a line chart that showed wind speed for the past hour and a CI forecast of wind speed over the next hour, and were asked to ``look at the wind speed forecast above and decide what to do with the electrical circuit'' (see~\cref{fig:teaser}, a). After the trials, participants took a Short Graph Literacy test~\cite{okan-sgl}, and answered demographic questions. All surveys are in the  supplemental materials.
\newline

\begin{figure}[t]
  \centering
\includegraphics[width=0.96\columnwidth, alt={A four-panel figure showing the high and low quantitative uncertainty conditions. The top row displays 50\% confidence intervals with the ensemble median, and the bottom row displays the corresponding generating ensemble forecasts.}]{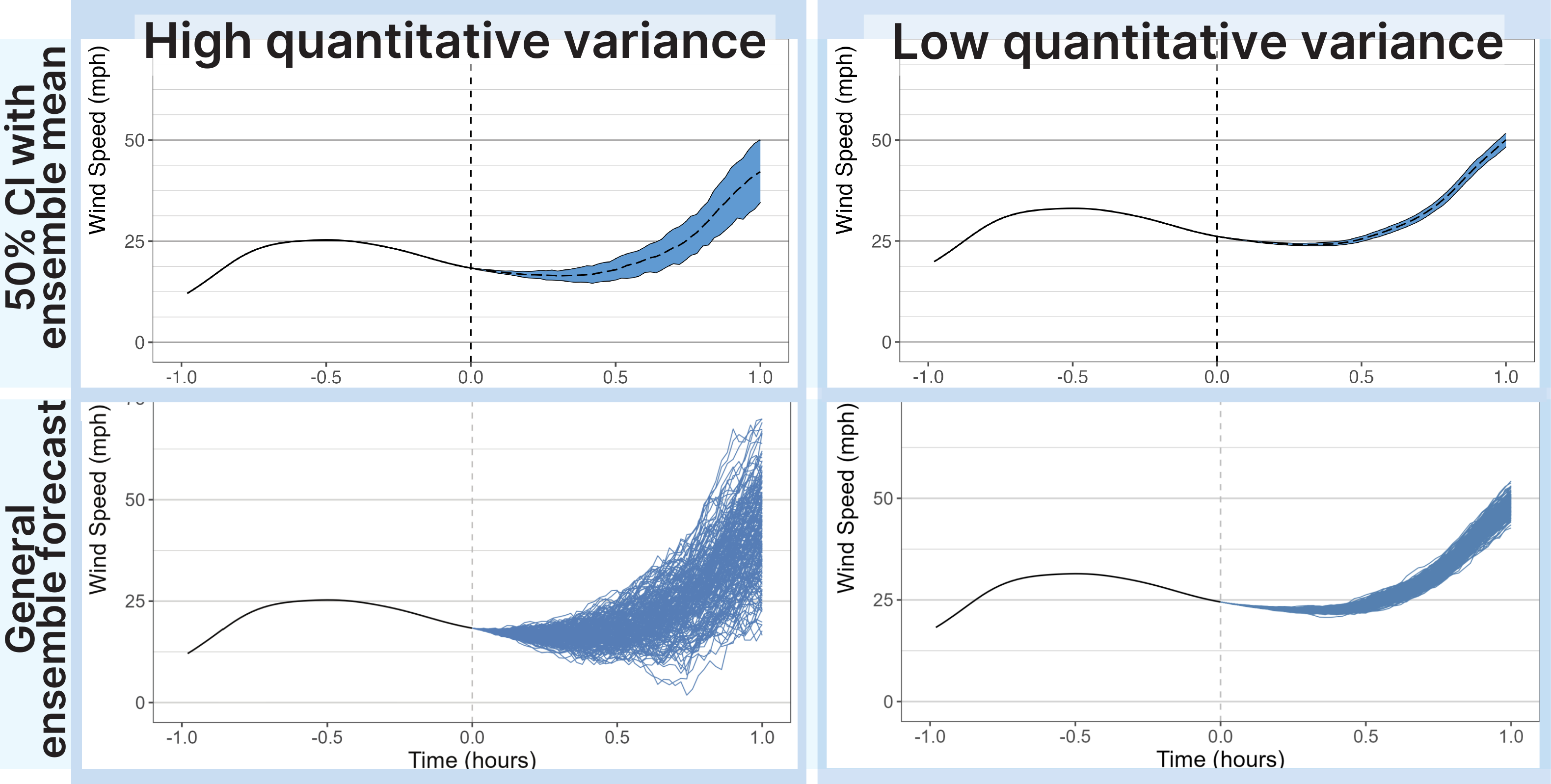}
  \caption{%
    Experiments' quantitative uncertainty conditions. Left: high quantitative uncertainty. Right: low quantitative uncertainty. Top: 50\% confidence interval with ensemble median. Bottom: corresponding generating ensemble forecast.%
  }
\label{fig:quant_uncertainty_levels}
\vspace{-1em}
\end{figure}

\noindent\textbf{Stimuli.} We generated all wind speed forecasts in R, producing six distinct shapes, that varied the trajectory of past and future forecasted wind data (\cref{fig:six_forecast_shapes}). These shapes added visual variety to our stimuli and ensured that our findings would be more generalizable across many scenarios. %
We created the forecasts by generating 200 wind speed plots using a Gaussian distribution. We varied the spread parameter of the forecasts for narrower (x1) or wider (x5) outcome ranges to create multiple levels of quantitative variance for our experiments (see \cref{fig:quant_uncertainty_levels}). %
We also varied the forecasts' level of risk by shifting entire plots along the y-axis, such that the percentage of forecasts that crossed the turn-off threshold at their maxima varied. We refer to this manipulation as the "\% Crossing" condition. We generated our final stimuli by plotting the 50\% CIs and mean wind speed for each set of forecasts.
\newline

\noindent\textbf{Reward Scheme.}
\label{sec:bonus}
Making a PSPS decision requires balancing risks and costs. %
To reflect this, and to create a fully-defined decision problem, %
we tied participants' decisions to bonus monetary incentives. %
We associated each forecast scenario with a predetermined outcome in which the ``actual'' wind speed crossed (or did not cross) a 50 mph turn-off threshold. These outcomes were aligned with the quantitative information displayed in visualization stimuli. For example, for the set of trials where stimuli indicate a 25\% probability of wind speed crossing the turn-off threshold, the predetermined outcomes that decided monetary rewards crossed the threshold 25\% of the time. For each trial in which a participant chose an action that was correct given the ``actual'' outcome, their bonus incremented by 0.05 USD. Each decision to turn off the circuit, cost 0.02 USD from their bonus to reflect the costs of real PSPS and to incentivize participants to leave the circuit on as often as possible. However, if a participant left the circuit on when the ``actual'' wind speed crossed the threshold, their bonus decremented 0.10 USD to reflect the higher costs of wildfire risk. To prevent learning effects, we did not show participants the outcome of each trial, and only informed them of their total bonus at the end of the experiment. However, before they started experimental trials, we informed participants about this cost/benefit structure, as well as the forecast generation technique, assumptions, appropriate interpretations, and sources of uncertainty to ensure they had sufficient information to make rational decisions~\cite{Hullman-2025-underspecified}.

Given the cost/benefit structure described above, the \textit{optimal} decision pattern was to leave the circuit on if $\leq$ 25\% of the wind speed forecast crossed the critical threshold (i.e., a 25\% chance of wind going too fast), and to turn it off $\geq$ 50\% of the forecast crossed the threshold. Note that when the turn-off forecast probability was 25\% or lower, the 50\% CI was either just below or just touching the threshold (i.e., 50 mph line) in the plot. If participants used a simple heuristic to turn off the circuit any time the CI went above the critical threshold, their responses would match the optimal pattern described above. We did not inform participants of this strategy, but past research suggests that 50\% CIs are the quantitative uncertainty visualization that best supports optimal decision making with this task~\cite{matzen2025preprint}.

\subsection{Experiment 1}
Experiment 1 investigated if communicating qualitative certainty about a quantitative time-series forecast impacts decision making. Past research on uncertainty communication for 1-dimensional data (e.g., probability distributions of a single random variable) suggests that showing qualitative and  quantitative uncertainty together can lead to more risk-averse decisions~\cite{padilla-2021-unc-about-unc}. Experiment 1 examines if these findings replicate in a different data context, namely for quantitative forecasts that are longitudinal, and therefore 2-dimensions.%

\subsubsection{Stimuli}
\label{sec:stim-exp1}

We created unique CI forecasts using the process described in \cref{sec:taskstim}. Then, we modified each forecast to have three different qualitative confidence levels---low, high, and absent. In the low and high confidence levels, text appeared below the forecast describing expert forecasters' confidence in the forecast (see~\cref{fig:teaser}, b). In the absent condition, the chart appeared alone, with no text statement.

\subsubsection{Experimental Design}
Experiment 1 utilized a 3 (\textit{Qual.~Confidence}: absent, low, high) x 6 (\textit{\% Crossing}: 0\%, 5\%, 25\%, 50\%, 75\%, 95\%) x 6 (\textit{Data Shape}: S1, S2, S3, S4, S5, S6) x 2 \textit{(Quant. Variance}: low, high) within-subjects design. This design resulted in 216 different stimuli which we distributed in a counterbalanced fashion to participants, each of whom viewed 36 stimuli that were evenly distributed across all variables. Our independent variables of interest were \textit{Qual.~Confidence} and \textit{\% Crossing}. We included multiple levels of \textit{Quant. Variance} and \textit{Data Shape} as repeated measures for better representation of the diversity of situations in which the PSPS task may occur. We did not have predictions for these variables and include them as covariates in our analysis. We collected participants’ binary decision to leave on or turn off the electrical circuit and use this decision as our dependent variable. Prior to running any of our experiments, we ran a series of pilot studies to
confirm question legibility, inform our hypotheses, and craft a power analysis to determine sample size.

\subsubsection{Hypotheses}
 Due to prior work that suggests that displaying qualitative uncertainty information with text alongside 1-dimensional probability distributions can impact decision making~\cite{padilla-2021-unc-about-unc}, we expected that adding the same information to time-series forecasts would have similar effects.
Thus, we hypothesized that \textbf{(H1)} participants’ likelihood of shutting off the electric grid would be more strongly affected by low qualitative confidence than by high qualitative confidence or no additional confidence information. We expected this effect to be most pronounced for forecasts near the optimal decision boundary (i.e., the 25\% and 50\% crossing levels), where the proportion of forecasts exceeding the turn-off threshold was just sufficient to shift the utility-maximizing decision from leaving the grid on (at 25\%) to turning it off (at 50\%).

\subsubsection{Participants}
\label{sec:participants-exp1}
Using pilot data and power simulations of our preregistered Bayesian model we determined a sample size of 300 participants would lead to a power level of > 0.8 for our main interaction of interest~\cite{goldfeld2021bayesianpower}. We recruited participants on Prolific.com who were at least 18 years old, fluent in English, residing in the United States, had a Prolific approval rate of 90\% or higher, and had not participated in our pilot studies. All participants completed the survey on a desktop or laptop computer. 

Although our tested task is inspired by decisions made by energy grid operators, we used a general-public, non-expert sample because our primary goal was to test the relative interpretive effects of qualitative confidence encodings, not to estimate expert decision maker performance. This is appropriate for a mechanism-oriented evaluation, where the aim is to assess how an encoding shifts responses under controlled conditions, rather than to characterize behavior in a professional target population~\cite{mook1983external}. At the same time, we do not assume that experts and non-experts share identical baseline behavior, as recent visualization research shows that participant populations can shape observed effects and should be described carefully~\cite{burns2023novices}. Therefore, we are careful to interpret our results primarily in terms of directional differences between visualization conditions, and caution our readers to do the same (see \cref{sec:limitations}). %

Of the 300 participants we recruited, two appeared to take our survey twice, and we removed all four of their responses for a total sample size of 296. Of this group, 159 identified as female, 131 as male, three as nonbinary, and three preferred not to report. In terms of age, 138 reported being between 18-30 years old, 127 were 31-50 years, 26 were 51-70 years, three were 71+, and two preferred not to report their age. The median graph literacy score was 2 of 4 (\textit{mean} = 1.89, \textit{SD} = 0.92). The median survey completion time was 13 minutes and 45 seconds. In all studies, participants were paid 2.80 USD, corresponding to roughly 12 USD per hour.
The median bonus per participant was 0.88 USD (\textit{SD} = 0.13).

\subsubsection{Procedure}

After indicating their consent, participants completed a Qualtrics~\cite{qualtrics} survey with the format detailed in \cref{sec:taskstim}. %
The survey and all stimuli are available in supplemental materials. 

\subsubsection{Analysis}
\label{sec:analysis-exp1}
We preregistered the following binomial Bayesian model:
\begin{equation}
\label{eq:exp1}
\begin{split}
    Binary\:Decision \sim \: Qual.\:\:Confidence \: \times \:\% \: Crossing \:+ \\Quant.\:\:Variance \:+ \:Forecast\:Shape\: + \\Graph\:\:Literacy \:+\:(ID \:\mid\:1)
\end{split}
\end{equation}
with uninformative priors centered at 0 with an SD of 2.5. We evaluate \textit{Binary Decision} as 0 if the circuit was left on, and 1 if a participant turned off the circuit. Because we fit this model on every question/answer combination, we include participant ID as a random intercept to account for participant-specific effects.

We implement this model in R using brms v 2.23.0~\cite{brms} and visualize its posteriors using tidybayes v 3.0.7~\cite{tidybayes}. We investigate the model's output through a visual analysis of its posteriors and by considering any effect coefficients that do not include zero in their 95\% credible interval to be meaningful, while avoiding a strictly dichotomous interpretation~\cite{kruschke2021bayesianreporting}.

\subsection{Experiment 2 \& 3}
Experiments 2 and 3 investigate visual methods for presenting qualitative confidence. In Experiment 2, we add glyphs to the textual caption from Experiment 1. In Experiment 3, we encode qualitative confidence directly on top of the forecast CI\rev{, instead of alongside it, a key functionality that is made possible by visualization}. Across both experiments, we test variations of blur, transparency, and color hue. We describe the motivation for these designs in Section~\ref{sec:stimuli-exp3}.

\begin{figure}[t]
  \centering
  \includegraphics[width=0.93\columnwidth, alt={Four vertical stacks of three circles that vary in how much they are blurred. Far left (uncertain-aligned): stack with most blurred circle at the top, followed by a medium blur circle, and a least blurred circle, with ``Uncertain'' corresponding to the top, blurred circle, and ``Certain'' corresponding to the bottom. Middle left (uncertain-opposite): stack with lease blurred circle at the top, followed by a medium blur circle, and a most blurred circle, with ``Uncertain'' corresponding to the top, least blurred circle, and ``Certain'' corresponding to the bottom, most blurred circle. Middle right (danger-aligned): stack with most blurred circle at the top, followed by a medium blur circle, and a least blurred circle, with ``High Danger'' corresponding to the top, most blurred circle, and ``Low Danger'' corresponding to the bottom, least blurred circle. Far right (danger-opposite): stack with lease blurred circle at the top, followed by a medium blur circle, and a most blurred circle, with ``High Danger'' corresponding to the top, least blurred circle, and ``Low Danger'' corresponding to the bottom, most blurred circle.  An example question text reads: ``Please rank how intuitively these visuals represent [condition] as shown above.'' with a scale from 1 to 7, wherein 1 is labeled ``illogical'' and 7 is labled ``logical''.}]{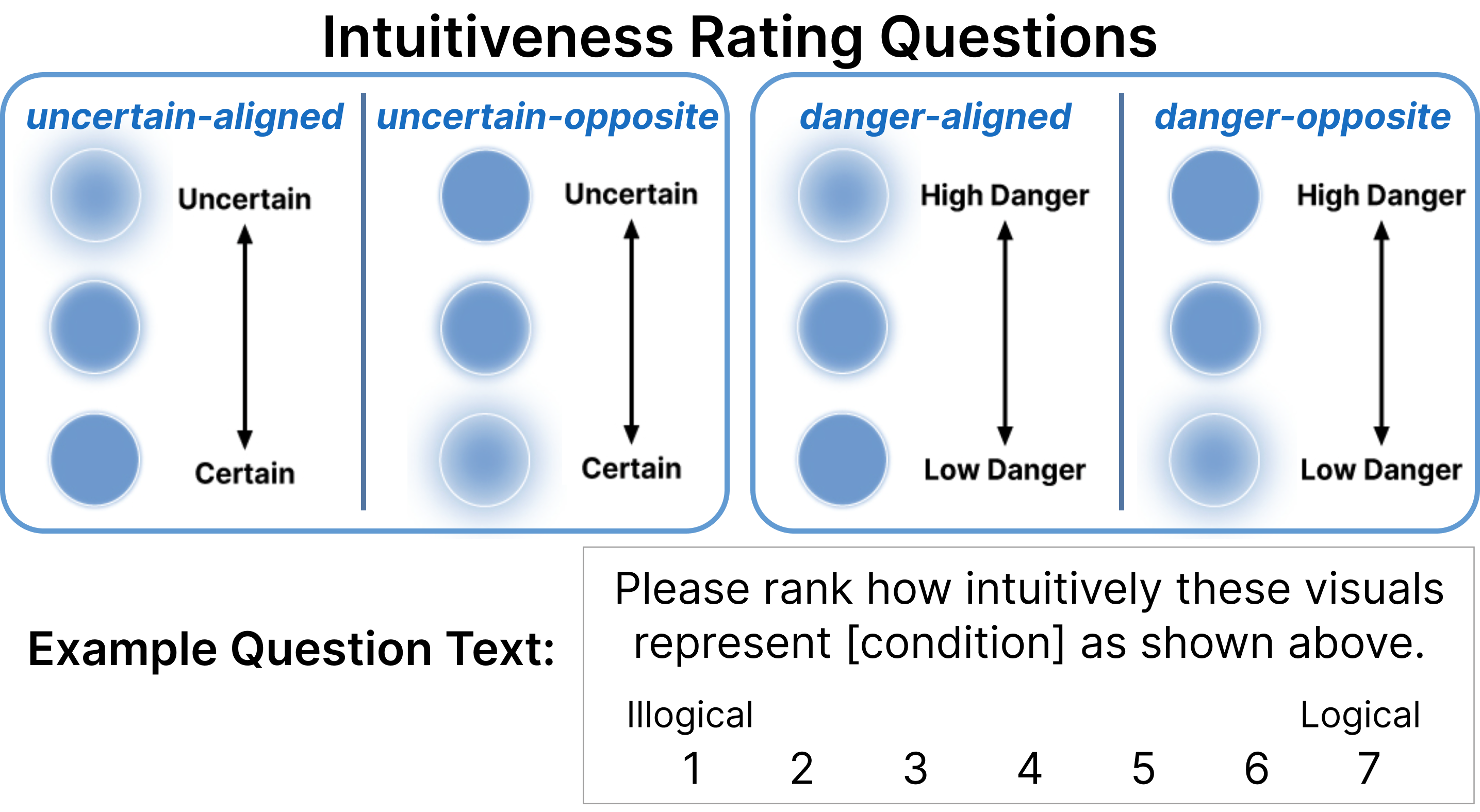}
  \caption{%
    Intuitiveness rating task~\cite{maceachren2012visual} for aligned and opposed mappings of uncertainty and danger using blur glyphs as an example condition.%
  }
  \label{fig:intuitiveness_ratings}
  \vspace{-1em}
\end{figure}

\subsubsection{Experimental Design}
\label{sec:experiment-design-exp2}
In both experiments, we maintained many of Experiment 1's design decisions, but replaced the \textit{Qual.~Confidence}'s absent level with a medium confidence level for further resolution of how changing forecaster confidence can impact behavior. We also added a third variable of interest, \textit{Confidence Encoding}, to evaluate techniques for visually representing qualitative confidence (e.g., blur, transparency, color hue) in a forecast. 

We only evaluated the middle four \textit{\% Crossing} levels, discarding the 0\% and 95\% levels from Experiment 1 because they showed rough behavioral consensus and thus were of less interest. Lastly, we shifted both experiments to a mixed design, using our new \textit{Confidence Encoding} variable to split our survey into a between-subjects design. 

These changes resulted in a  4 (\textit{Confidence Encoding}: text, blur, transparency, color hue) x 3 (\textit{Qual.~Confidence}: low, medium, 
high) x 4 \rev{(}\% Crossing: 5\%, 25\%, 50\%, 75\%) x 6 (\textit{Data Shape}: S1, S2, S3, S4, S5, S6) x
2 (\textit{Quant. Variance}: low, high) mixed design for Experiments 2 and 3. We created 576 stimuli for each experiment and distributed them \rev{using a Latin square design}, such that each participant viewed 24 stimuli that were evenly distributed across all within-subject variables.

Both experiments' independent variables of interest were \textit{Confidence Encoding}, \textit{Qual.~Confidence}, and \textit{\% Crossing}. We included the \textit{Quant. Variance} and \textit{Data Shape} as repeated measures, for the same reason as in Experiment 1. Both experiments tested an identical task to Experiment 1 (see \cref{sec:taskstim}). 
Prior to running both experiments, we ran pilot studies to inform our hypotheses and power analysis.

\subsubsection{Hypotheses}
As in Experiment 1, we expected that showing qualitative uncertainty alongside a wind speed forecast would impact participants' decision making because of past findings that indicate similar effects~\cite{padilla-2021-unc-about-unc}. We thought this would especially be the case as forecasts neared their optimal decision boundary and the obviousness of a clearly rational choice waned.
Thus, we preregistered the hypothesis that visualizations that express qualitative confidence using glyphs (\textbf{H2a}) and those that encode confidence directly on top of forecast CIs (\textbf{H3a}) would exhibit a \textit{Qual.~Confidence} × \textit{\% Crossing} interaction. In other words, we hypothesized we would see an effect in which low qualitative confidence led to a meaningfully different likelihood of turning off the circuit in comparison to high qualitative confidence, and that the \textit{\% Crossing} condition would meaningfully change this difference. We did not hypothesize about the effects of medium confidence.

Past research also suggests that varying quantitative uncertainty representation can impact optimal decision making in PSPS and other contexts\cite{matzen2025preprint, kay2016ish, kale-visual-reasoning-strategies-2021}, so we expected similar effects to stem from the variation of qualitative uncertainty encodings.
Thus, we also preregistered a hypothesized three-way \textit{Qual.~Confidence} × \textit{\% Crossing} × \textit{Confidence Encoding} interaction. %
Specifically, that the \textit{Qual.~Confidence} × \textit{\% Crossing} effect described above would vary as a function of forecasts' \textit{Confidence Encoding} (blur, transparency, and color hue glyphs, and text for \textbf{H2b}; overlaid blur, transparency, and color hue encodings, and text for \textbf{H3b}).

\subsubsection{Stimuli}
\label{sec:stimuli-exp3}
We generated all of our windspeed forecast stimuli using the methods described in \cref{sec:taskstim}. We derived the designs for Experiment 2's qualitative confidence glyphs and Experiment 3's overlaid encodings (see~\cref{fig:teaser}, f through k.) from the series of visual channels investigated by MacEachren et al. in their work on the visual semiotics of uncertainty representations~\cite{maceachren2012visual}. We selected three channels investigated by MacEachren et al.: \textbf{blur} (also referred to as ``fuzziness''), which exhibited the strongest ratings of representing uncertainty intuitively; \textbf{transparency}, which exhibited moderate ratings of intuitiveness; and \textbf{color hue}, which exhibited low ratings of intuitiveness in representing uncertainty. Notably, blur and transparency had directionality associated with their intuitiveness (e.g., high blur and transparency more intuitively represented      ``uncertainty'' than low blur and transparency), whereas color hue did not. We designed our glyphs and overlaid encodings to align with this directionality, when present. For color hue, we selected a traffic-light red-yellow-green color scheme commonly used in risk communication and adjusted the lightness of each color to maintain grayscale differentiability, ensuring the hues are distinguishable regardless of color-vision deficiency.

All of Experiment 3's stimuli were generated using the R scripts that generated wind speed forecasts for Experiments 1 and 2. We modified these scripts to encode qualitative confidence into the forecasts' confidence interval, as shown in ~\cref{fig:teaser}, i-k. 
Experiment 3's color hue stimuli were modified to have the same color-vision-deficient-friendly, traffic-light color scheme as used in Experiment 2's color hue glyphs.

\begin{figure}[h!]
  \begin{center}
\includegraphics[width=0.45\textwidth, alt = {Two confidence intervals on line chart. The left interval is a concetrated blue with little blur, and the right interval is more blurred which makes it a less concentrated, lighter blue color.}]{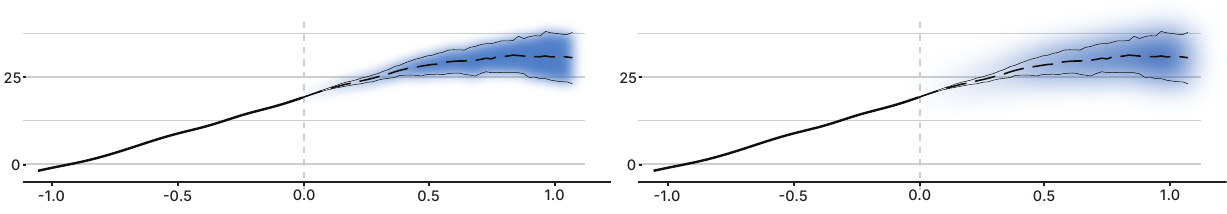}
  \end{center}
  \vspace{-1em}
  \caption{Two blur levels illustrate that greater blur makes a CI appear wider.}
  \label{fig:blur}
  \vspace{-1em}
\end{figure}

Experiment 3's blur condition proved more challenging. Because blurring a visual object impacts its perceived size and position~\cite{held-2010-blur-perceived-size, schroeger-2021-blur-effects-on-precision}, we are unable to create overlaid blur stimuli that mirror Experiment 2's glyphs without modifying the placement of the CIs' outer edge given the level of qualitative confidence represented (see \cref{fig:blur}). Although modifying distributions to better align readers with rational decisions is an option~\cite{yang2023subjective}, we abstained from doing so, because the degree to which low, medium, and high qualitative confidence should rationally change a PSPS decision is an unanswered question outside the scope of this experiment. Instead, as this experiment is motivated by the intuitiveness of blur in communicating uncertainty, we developed a blurred-stroke design to maintain a blurred aesthetic without altering CI spread.
This design adds a gray, blurred stroke around the perimeter of each forecast CI, varying the stroke's darkness by certainty level. We designed the gray, blurred stroke to convey the concept of ``fog'' or ``smoke,'' common metaphors that align with uncertainty~\cite{MacEachren-2005-geospatial-unc, Pokojna-2025-vismetaphor}. Thus, we represented high qualitative confidence with a light gray ``fog'', medium confidence with a darker gray fog, and low confidence with a black fog (see~\cref{fig:teaser}, i).  Although this fog modification changes the visual edge of the CIs altogether, the expanded edge does not vary across confidence levels, nor does it obstruct or interfere with the inner CI edge that encodes quantitative variability. %
All stimuli\rev{, and more details of their generation,} are available in supplemental materials.

\subsubsection{Participants}
As in Experiment 1, for Experiments 2 and 3 we collected pilot data to inform a power analysis of our preregistered Bayesian model~\cite{goldfeld2021bayesianpower}. This analysis indicated a sample size of 75 participants per between-subject group for a power of $\geq$ 0.7 for our interaction of interest. In both experiments, we rounded up to 78 participants so as to be evenly divisible by our six counterbalanced subgroups. We maintained all the same recruiting criteria, methods, and payment as in Experiment 1. 

For Experiment 2, due to some variability in Qualtric's randomizer function, we collected 79 participants for the text and color hue glyph conditions, and 78 participants for the other conditions \rev{(total \textit{n} = 314)}. In the experiment, 167 identified as female, 136 as male, 9 as nonbinary, and two preferred not to report. 152 were 18-30 years old, 135 were 31-50 years, 25 were 51-70 years, zero were 71+, and two preferred not to report their age. 5 participants identified as colorblind. The median graph literacy score was 2 of 4 (\textit{mean} = 1.89, \textit{SD} = 0.89). The median survey completion time was 13 minutes and 35 seconds. The median bonus per participant was 0.50 USD (\textit{mean} = 0.45, \textit{SD} = 0.22).

For Experiment 3, we recruited 78 participants per condition \rev{(\textit{n} = 312)}.
Overall, 157 participants identified as female, 142 as male, 11 as nonbinary, and two \rev{did} not report. 151 were 18-30 years old, 131 were 31-50 years, 28 were 51-70 years, one was 71+, and one \rev{did} not report their age. 6 participants identified as colorblind. The median graph literacy score was 2 of 4 (\textit{mean} = 1.94, \textit{SD} = 0.85). The median survey completion time was 13 minutes and 32 seconds.  The median bonus per participant was 0.51 USD (\textit{mean} = 0.47, \textit{SD} = 0.21).

We used the Experiment 2's text condition as the comparison baseline for Experiments 2 and 3. This was possible because \textit{Confidence Encoding} was manipulated between subjects in both experiments.%

\subsubsection{Procedure}
\label{sec:procedure-exp2}
Our procedures for Experiments 2 and 3 were similar to those for Experiment 1. We used the same attention and comprehension checks, and modified the instructions to explain how to read the qualitative confidence encoding for each between-subjects group. We also added four follow-up questions in the style of MacEachren et al.'s intuitiveness experiment~\cite{maceachren2012visual}. %
The intuitiveness questions showed participants an ordered set of the three levels of encoding condition they saw in the wind speed forecasts, with the uncertainty mapping used in the experiment, and then with an inverted mapping (see \cref{fig:intuitiveness_ratings}).
We asked participants to rate how intuitively the glyphs represented not only uncertainty, but also danger, for a total of four intuitiveness questions. %

We included these questions because we were interested in how our participants viewed the meaning of our tested confidence encodings. We investigated danger, as well as uncertainty, because the red-yellow-green scheme in our color conditions is commonly associated with danger, %
and could shed light on potential differences in decision making. We did not preregister hypotheses about these questions' results.

\subsubsection{Analysis}
\label{sec:AnalysisExp2}

To analyze qualitative confidence encodings' impact on decision making we preregistered a binomial Bayesian model with a three-way interaction between \textit{Confidence Encoding}, \textit{Qual.~Confidence} and \textit{\%~Crossing}. Prior to analysing results, we compared this model against a model containing the three corresponding two-way interactions to determine which better fit our data. We conducted this comparison using leave-one-out cross-validation with the loo package v2.8.0 in R~\cite{vehtari-loo-paper-2017, loo-r}.
We found the model with only two-way interactions produced a better fit (expected log predictive density difference = -11.5 with standard error of 4.0, and -10.1 with error of 4.8 for Experiment 2 and 3, respectively).  
Thus, we report results from the following model: 
\begin{equation}
\label{eq:exp2}
\begin{aligned}
    Binary\:Decision \sim \: Qual.\:\:Confidence \: \times \:\% \: Crossing \:+\\ Qual.\:\:Confidence \: \times \:Confidence \:Encoding \:+ \\ Confidence \:Encoding \: \times \:\% \: Crossing \:+\\ \: Quant\:\:Variance + Forecast\:Shape\: + \\\:Graph\:Literacy \:+\:(ID \:| \:1)
\end{aligned}
\end{equation}
with uninformative priors centered at 0 with an SD of 2.5. We implement this model using the same software, and interpret it using the same methods, as in Experiment 1 (see \cref{sec:analysis-exp1}). 

To analyze the intuitiveness questions, we fit an exploratory Bayesian cumulative logistic regression model using the following equation:
\begin{equation}
\label{eq:intuit}
\begin{aligned}
   Intuitiv\rev{e}ness\:Rating \sim \: \\Confidence\:Encoding \: \times \:Intuitiveness \: Question \:+\:(ID \:| \:1)
\end{aligned}
\end{equation}
\noindent where \textit{Intuitiveness Rating} was an ordinal factor ranging from 1 to 7 and \textit{Intuitiveness Question} reflected each of the four questions we asked (see \cref{sec:procedure-exp2}). As the procedures for all between-subjects groups in Experiment 2 and 3 were very similar, we ran the exploratory analysis of \textit{Confidence Encoding} conditions in Experiment 2 and 3 together. %

\section{Results}
We discuss both \rev{preregistered and exploratory} experimental results. %
We consider \textit{meaningful effects} to be those with a predicted coefficient whose 95\% credible interval (CrI) excludes zero.%

For all three experiments, we preregistered the exclusion of participants who failed to answer a simple attention check exactly as instructed, if they meaningfully skewed our results. \rev{All participants, including those who failed this initial check, then completed comprehension checks throughout the instructions to ensure they understood the premise and goals of the experimental task.} We conducted a sensitivity analysis for each experiment, comparing results from its entire sample with those from only participants who passed the \rev{initial} check. Experiment 1 and 2 showed no meaningful differences between these two groups. Experiment 3 had differences in the meaningfulness of three predicted coefficients, although all were relatively small (mean differences < 2.5\%) and not readily apparent in a visual analysis of the posteriors.
For a larger sample size and more conservative analysis, we present the results from each experiments' full samples.
All sensitivity analyses %
are available in supplemental materials.

\subsection{Experiment 1}
\begin{figure}[t]
 \includegraphics[width=\linewidth, alt={Posterior density plots show the probability of shutting off the circuit (0\% - 100\% along the x-axis), for each of six Percent Crossing conditions (0\% crossing, 5\% crossing, 25\% crossing, 50\% crossing, 75\% crossing, 95\% crossing along the y-axis). Three posteriors per percent crossing conditions are shown: one corresponding to stimuli with no/absent qualitative confidence, one corresponding to stimuli with low confidence, and one to stimuli with high confidence. Along the right side of the figure, brackets display meaningful differences of differences between qualitative confidence pairs across different percent crossings. Posteriors generally form a pattern from left to right: the lower the percent crossing (down on y-axis), the lower the probability of shutting off the circuit (left on the x-axis), with a gap betwen the 25\% and 50\% crossing conditions, which makes the shape of posteriors approximate a tan line shape.}]{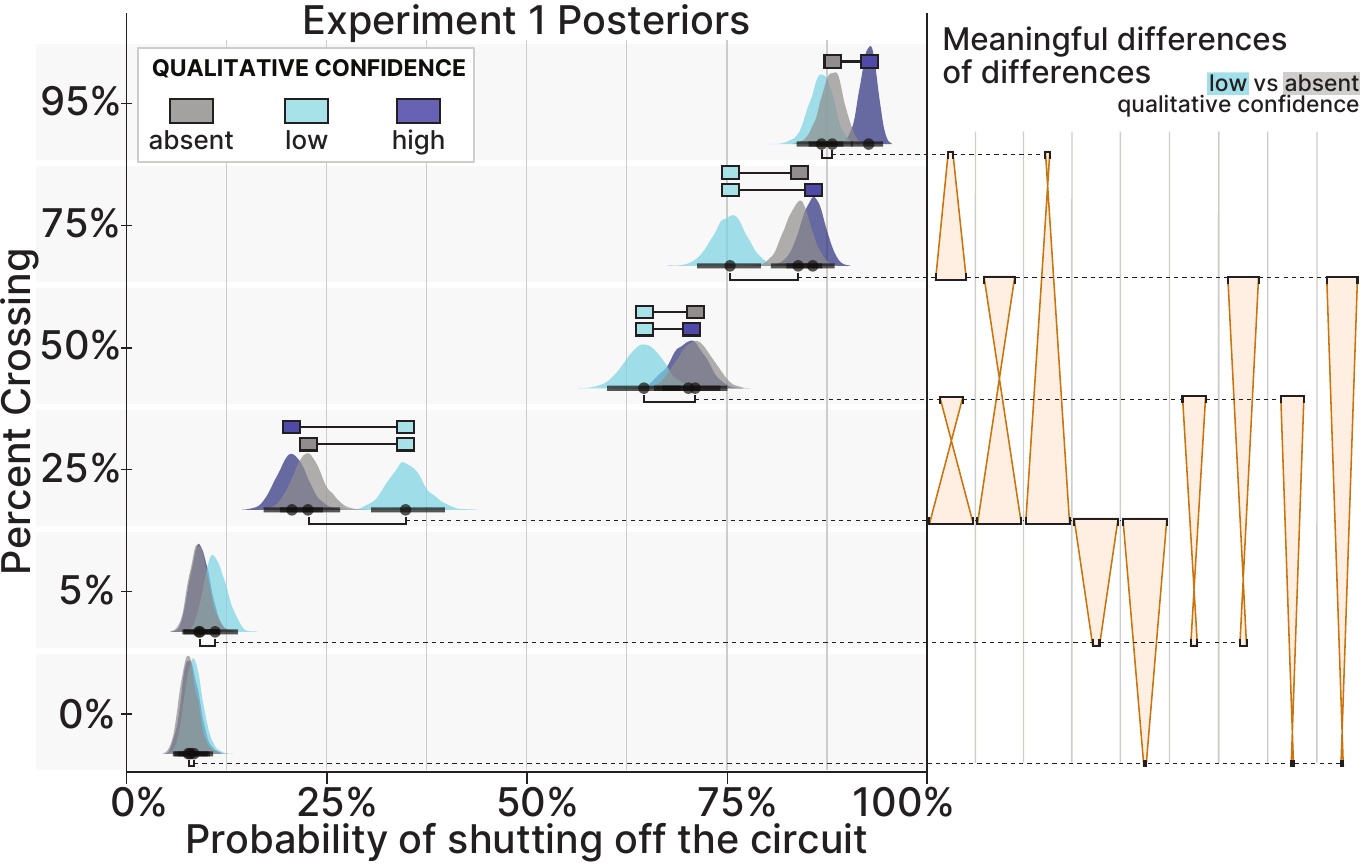}
  \caption{%
  	Posteriors from \cref{eq:exp1}. %
\includegraphics[height=0.5em, alt={connected rectangle pairs}]{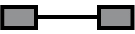} show meaningful differences.\\ \includegraphics[height=0.9em, alt={orange brackets}]{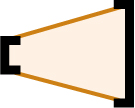} show meaningful second-order interactions between low and absent qualitative confidence and different encodings. Crossed \includegraphics[height=0.9em, alt={orange brackets}]{figs/exp1-inteffect-annot.jpg} indicate reordering of high and low posteriors. Not all meaningful second-order interactions are shown; see \cref{tab:exp1-ests} in Appendix.}
  \label{fig:exp1-results}
  \vspace{-1em}
\end{figure}
\subsubsection{Binary Decision}
We investigate whether the binary decision to turn off an electrical grid changes as the percent of the wind speed forecast that crosses a turn-off threshold changes, and as low or high qualitative confidence is displayed alongside forecasts. We hypothesized that showing low confidence forecasts would lead to a meaningfully different likelihood of turn-off
than forecasts absent of, and with high confidence, captions. We hypothesized that this effect would be particularly present at \textit{\% Crossing} levels near the optimal decision boundary, where a rational agent would switch between leaving the circuit on and turning it off. 

\begin{figure}[h!]
    \centering
\includegraphics[width=0.42\textwidth, alt={Posteriors show low probability of shutting off the circuit at low percent crossing, and higher probability of shutting of the circuit at higher percent crossings. Glyph transparency exhibits meaningful difference of differences with other glyph/text conditions across all percent crossings.}]{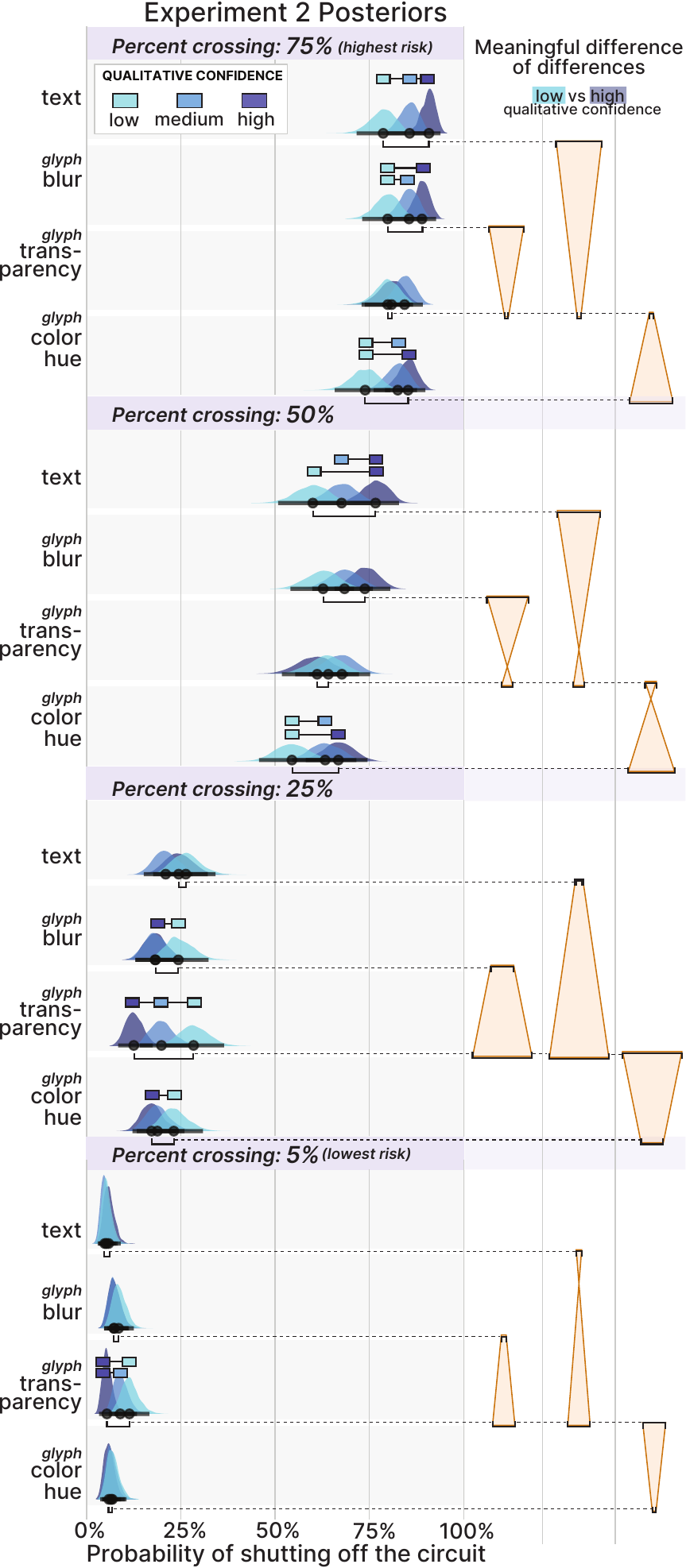}
    \caption{\cref{eq:exp2} posteriors modeled on Experiment 2 data. %
\includegraphics[height=0.5em, alt={connected rectangle pairs}]{figs/expintuiti-maineffect-annot.jpg} show meaningful differences. \includegraphics[height=0.9em, alt={orange brackets}]{figs/exp1-inteffect-annot.jpg} show meaningful second-order interactions between high and low qualitative confidence and different encodings. Crossed \includegraphics[height=0.9em, alt={orange brackets}]{figs/exp1-inteffect-annot.jpg} indicate reordering of high and low posteriors. Not all meaningful second-order interactions are shown; see \cref{tab:exp2-ests} in Appendix.}
    \label{fig:exp2-results}
    \vspace{-1em}
\end{figure}

\noindent\textbf{Low Qualitative Confidence.}
We investigate this hypothesis using the Bayesian model \cref{eq:exp1} with \textit{Qual.~Confidence}'s referent set to absent, and then to low. Next, we vary the referent for \textit{\% Crossing} across all levels to see how the two variables interact. %
This analysis reveals that adding a low qualitative confidence caption to a forecast meaningfully changes the likelihood of participants shutting off the circuit at the 25\%, 50\% and 75\% crossing levels, \textbf{\rev{providing strong support for} H1}. These effects are indicated with~\includegraphics[height=0.6em, alt={connected light blue and gray rectangle pairs}]{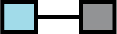} in \cref{fig:exp1-results}. 

We also find that the difference between absent and low confidence information meaningfully changes across many \textit{\% Crossing} levels. These effects are indicated with~\includegraphics[height=0.9em, alt={orange brackets}]{figs/exp1-inteffect-annot.jpg}\rev{, wherein the difference between the absent-low confidence pair at one end is meaningfully different from the difference between the absent-low confidence pair at the other end.} Most interestingly, this relationship at \rev{the} 25\% crossing differs meaningfully from that at 50\% and 75\% crossings.
Visual investigation of posteriors reveals that low confidence \textit{increases} likelihood of turn-off compared to absent and high confidence at the 25\% crossing level, but \textit{decreases} this likelihood at the 50\% and 75\% levels.%

\noindent\textbf{High Qualitative Confidence.}
For the majority of \textit{\% Crossing} levels, adding high qualitative confidence text to a forecast does not meaningfully alter participants' likelihood of turning off the circuit. %
The 95\% crossing level is the only exception. At 95\% crossing, adding a high qualitative confidence caption meaningfully shifts participants towards turn-off, more so than absent or low confidence captions. The top row of \cref{fig:exp1-results} shows that, although this shift is meaningful, all three \textit{Qual.~Confidence} levels are tightly clustered around 80-95\% likelihood of turn-off. Note that high confidence had meaningful second-order effects that are not annotated in \cref{fig:exp1-results}; see \cref{tab:exp1-ests} in the Appendix.%

\begin{figure*}
    \centering
\includegraphics[width=0.9\linewidth, alt={posteriors show a range of alignments between all 7 encoding conditions and uncertain-aligned, uncertain-opposite, danger-aligned, and danger-opposite scales. Each scale ranges from 1 (illogical) to 7 (logical).}]{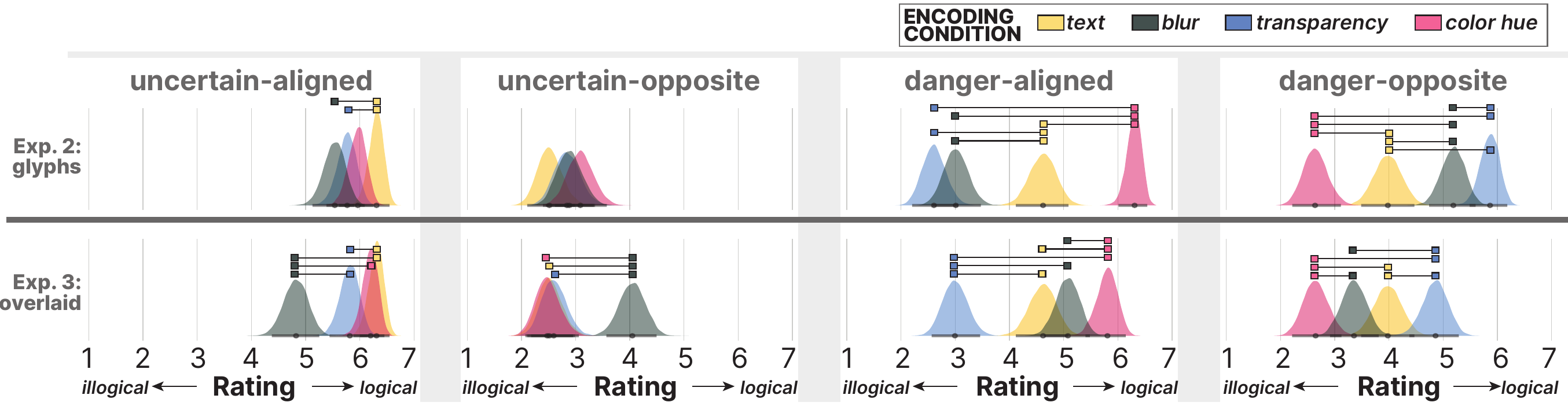}
\vspace{-2mm}
    \caption{Intuitiveness rating posteriors modeled from \cref{eq:intuit} on Experiment 2 data (top) and Experiment 3 data (bottom). Columns show ratings for each question in \cref{fig:intuitiveness_ratings} and colored posteriors are divide by \textit{Confidence Encoding}. %
\includegraphics[height=0.5em, alt={connnected rectangle pair}]{figs/expintuiti-maineffect-annot.jpg}~denotes meaningful differences between posteriors.}
    \label{fig:intuit-results}
    \vspace{-1em}
\end{figure*}

\subsubsection{Experiment 1 Results Summary}
Overall, Experiment 1 showed that qualitative confidence influenced decisions, with its effects depending strongly on the underlying level of forecast risk. Low confidence meaningfully shifted turn-off decisions at the 25\%, 50\%, and 75\% crossing levels, increasing turn-off likelihood at 25\% crossing but decreasing it at 50\% and 75\% crossing. In other words, participants were more likely to choose the cautious option when forecasts were just below the optimal turn-off threshold, but less likely to commit to turn-off when forecasts were just above it. In contrast, high confidence had relatively little effect across most crossing levels, except for 95\% crossing, where it increased turn-off likelihood. Together, these results suggest that low expert confidence made participants less certain about a forecast's warranted action.

\subsection{Experiment 2}
Experiment 1 revealed that qualitative forecaster confidence shapes decisions even when quantitative uncertainty is available, with the strongest effects appearing near decision boundaries. Experiment 2 extends this result by investigating if communicating confidence through labels with glyphs (see \cref{fig:teaser}, f-h) has any impact on decisions.

\subsubsection{Binary Decision}
\label{sec:binary-dec-exp2}
We predicted Experiment 2 would replicate Experiment 1's finding that low and high qualitative confidence produce different turn-off decisions, especially across levels of forecast risk (H2a). We further predicted that this effect would be different for text and glyph-based expressions of forecaster confidence~(H2b).
As shown in~\cref{fig:exp2-results}, Experiment~2's posteriors largely mirrored those of Experiment~1. Participants were more likely to turn off the circuit as \textit{\% Crossing} increased. Also, low confidence was generally associated with a higher likelihood of turn-offs than high confidence at 25\% crossing, %
and a lower likelihood at 50\% and 75\% crossings. %
\rev{Thus,} we \rev{find \textbf{moderate support for H2a}}.%

Experiment~2 also revealed that the medium qualitative confidence posteriors are not consistent in their relationship to low and high confidence posteriors. %
Meaningful differences in the likelihood of turn-off between medium and high qualitative confidence at the 5\% or 25\% crossing levels were rare, and only sometimes appeared at higher crossing levels.
This pattern suggests a non-linear characterization of qualitative confidence, although future work must investigate this effect at a higher resolution before extensible conclusions are possible.

With Experiment~2, we also assessed if glyphs could communicate qualitative confidence. As described in Section~\ref{sec:AnalysisExp2}, \cref{eq:exp2}, which has three two-way interactions, better fit our data than its counterpart with a three-way interaction between \textit{Qual.~Confidence}, \textit{Confidence Encoding}, and \textit{\% Crossing}. This indicates that we found no evidence that the relationship between \textit{Qual.~Confidence} and \textit{Confidence Encoding} varied across \textit{\% Crossing} levels, which \textbf{\rev{provides no support for} H2b}. 

Interestingly, \rev{exploratory} follow-up contrasts across \textit{\% Crossing} levels reveal \rev{a potential,} more nuanced pattern for transparency glyphs. %
Transparency glyphs meaningfully changed the deltas between low and high confidences' effect, but not in a uniform way. As shown via the transparency glyph ~\includegraphics[height=0.9em, alt={orange brackets}]{figs/exp1-inteffect-annot.jpg}~s' larger edge in the bottom \rev{half} of \cref{fig:exp2-results}, at 5\% and 25\% crossings transparency glyph's high-versus-low confidence contrast was meaningfully larger than other conditions' contrasts. %
 However, as transparency ~\includegraphics[height=0.9em,alt={orange brackets}]{figs/exp1-inteffect-annot.jpg}~s' smaller edges in the top two \rev{half} indicate, this effect was in the \textit{opposite} direction for 50\% and 75\% crossings%
. The estimates and credible intervals for all effects are available in \cref{tab:exp2-ests} in the Appendix. %
Across these comparisons, other \textit{Confidence Encoding} levels did not meaningfully differ.

\subsubsection{Confidence Encoding Intuitiveness}

We \rev{also completed an exploratory analysis of} how intuitively Experiment 2's \textit{Confidence Encodings} communicated the concepts of uncertainty and danger %
(questions in \cref{fig:intuitiveness_ratings}). Experiment 2's intuitiveness posteriors are in the top row of \cref{fig:intuit-results}. The uncertainty-aligned results indicate that text was rated as communicating uncertainty more intuitively than blur %
and transparency glyphs%
, but not color hue glyphs.%

The danger-aligned results show that color hue glyphs are much more intuitively aligned with danger than all other encodings. %
Blur and transparency glyphs showed relatively poor alignment with danger, as reflected by low ratings for danger-aligned mappings and high ratings for danger-opposite mappings, despite rating as highly uncertain-aligned. This pattern suggests an inverse relationship, where blur and transparency glyphs naturally communicate uncertainty, but not danger.

\subsubsection{Experiment 2 Results Summary}

In summary, Experiment 2 broadly replicated the pattern observed in Experiment 1, and we found no overall evidence that glyph-based encodings increased the effect of qualitative confidence relative to text, transparency glyphs \rev{being} the only exception. In comparison to text, these glyphs amplified confidence differences at lower crossing levels but reduced them at higher crossing levels, suggesting that \rev{they may have} alter\rev{ed} how participants incorporated qualitative confidence in a non-uniform manner. Blur and color hue glyphs did not change participants’ behavior more than text, suggest\rev{ing they} may serve as \rev{comparable} visual alternatives \rev{to text.}%

Experiment 2's intuitiveness ratings also revealed useful \rev{evidence}. Text was judged most intuitive for communicating uncertainty, color was most strongly aligned with danger, and blur and transparency appeared to communicate uncertainty more naturally than danger.

\begin{figure}[t!]
    \centering
    \includegraphics[width=0.40\textwidth, alt={posteriors show low probability of shutting off the circuit at low percent crossing, and higher probability of shutting of the circuit at higher percent crossings. Overlaid blur exhibits meaningful difference of differences with other overlaid transparency and text conditions across all percent crossings.}]{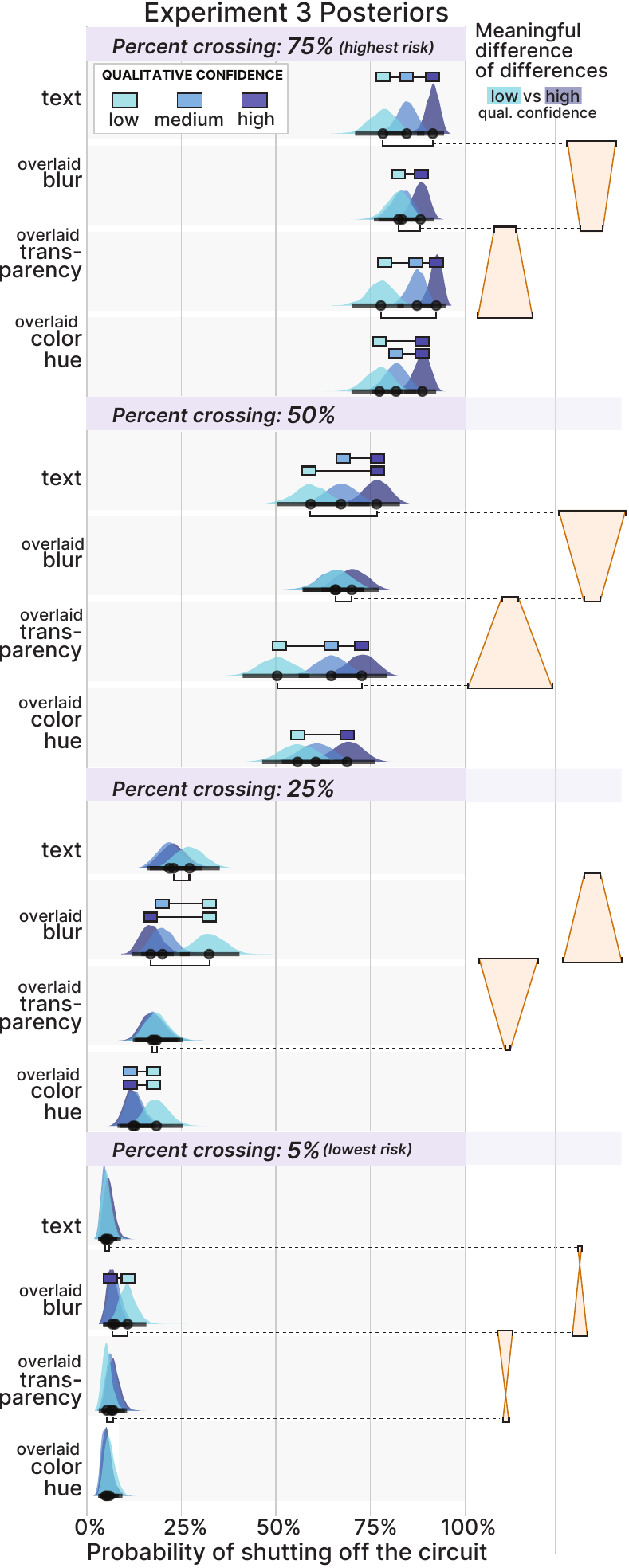}
    \vspace{-2mm}
    \caption{\cref{eq:exp2} posteriors  modeled on Experiment 3 data. %
\includegraphics[height=0.5em, alt={connected rectangle pairs}]{figs/expintuiti-maineffect-annot.jpg} show meaningful differences. \includegraphics[height=0.9em, alt={orange brackets}]{figs/exp1-inteffect-annot.jpg} show meaningful second-order interactions between high and low qualitative confidence and different encodings. Crossed \includegraphics[height=0.9em, alt={orange brackets}]{figs/exp1-inteffect-annot.jpg} indicate reordering of high and low posteriors. Not all meaningful second-order interactions are shown; see \cref{tab:exp3-ests} in Appendix.}
    \label{fig:exp3-results}
    \vspace{-1em}
\end{figure}
 
\subsection{Experiment 3}
Experiment 3 extends \rev{Experiment 2's results} by exploring if qualitative confidence can be communicated \textit{only visually} by encoding confidence level on top of forecast CIs (see~\cref{fig:teaser}, i-k), and if these overlaid representations alter decision-making differently than text.

\subsubsection{Binary Decision}
For Experiment 3, we made the same core predictions as in Experiment 2, but for overlaid, rather than glyph, encodings. We predicted that low and high qualitative confidence would produce different decisions, and that this difference would vary across \textit{\% Crossings}~(H3a). We further predicted that these effects would be different across text and overlaid \textit{Confidence Encoding} levels~(H3b).
We found evidence of a \textit{Qual.~Confidence} $\times$ \textit{\% Crossing} interaction in Experiment~3, providing \textbf{\rev{moderate support for} H3a}.
As in Experiment~2, the two-way interaction model in \cref{eq:exp2} fit the data better than its three-way interaction counterpart, so we \textbf{do \rev{not find evidence to support} H3b}. 

However, \rev{an exploratory analysis using \cref{eq:exp2}} reveals a pattern in Experiment~3 that was similar to transparency's interactions in Experiment~2, but for overlaid blur. Follow-up contrasts across the \textit{\% Crossing} levels in \cref{fig:exp3-results} reveal the same relationship inversion between blur's high-v-low contrasts and text's and transparency's. In the bottom \rev{half} of \cref{fig:exp3-results}, which reflect\rev{s} 5\% and 25\% crossings, overlaid blur's high-vs-low contrast is meaningfully larger than text's and transparency's, as indicated by the \includegraphics[height=0.9em, alt={orange bracket}]{figs/exp1-inteffect-annot.jpg}~\rev{s' larger blur edge}. However, in the top \rev{half, which shows} 50\% and 75\% crossings, the  \includegraphics[height=0.9em, alt={orange bracket}]{figs/exp1-inteffect-annot.jpg}~s' blur edges are smaller than their opposing edge, indicating blur's high-vs-low contrasts are meaningfully \textit{smaller} than text's and transparency's. %
We report all estimates and credible intervals in \cref{tab:exp3-ests} in the Appendix.

Taken together, these results \rev{provide moderate evidence} that overlaid blur amplified \rev{differences in decision making invoked} by high and low qualitative confidence at lower crossing levels, but reduced \rev{these differences} at higher crossing levels.

\subsubsection{Confidence Encoding Intuitiveness}

\rev{An exploratory analysis of} Experiment \rev{2 and Experiment 3's intuitiveness ratings revealed %
that} overlaid blur was rated as meaningfully less intuitive in communicating uncertainty than blur glyphs (\cref{fig:intuit-results}).  %
This change did not occur for any other \textit{Confidence Encoding} representations and could provide insight into the more cautious behavior \rev{found} in overlaid blur's decision trials. 
\rev{Inversely}, overlaid blur communicated danger more logically than blur glyphs. %
\rev{These} differences indicate that the overlaid blur design was not semantically consistent with its glyph counterpart\rev{, which }%
highlights the difficulty of translating blur's semantic affordances to line chart CIs.%

Otherwise, \textit{Confidence Encodings}' ratings were relatively stable.
There was a small but meaningful decrease between color hue's overlaid and glyph danger-aligned rating; the glyphs communicated danger more intuitively. The same was true for transparency overlay and glyphs, but in the danger-opposite question.

 \subsubsection{Experiment 3 Results Summary}
 Across Experiments 2 and 3, an interesting pattern emerged. Color hue encodings \rev{altered decision making in a fashion} similar to text in both experiments, suggesting color may be the most stable of the visual alternatives to textual confidence, whether implemented as a glyph or as an overlay. Blur glyphs and overlaid transparency also behaved similarly to text, \rev{across the 2-way interactions we analyzed}. 
 However, transparency glyphs and overlaid blur diverged from this overall effect pattern. %
 At 5\% and 25\% crossing, using these encodings to show low qualitative confidence produced more cautious decisions than other encodings, increasing the likelihood of grid turn-off.
 At 50\% and 75\% crossing, however, both encodings' high and low conditions were less likely to lead to different turn-off rates than other encodings. Overlaid blur's intuitiveness ratings provide a possible explanation for its pattern. Whereas most encodings were rated similarly across glyph and overlay formats, blur \rev{was not}. Overlaid blur was less aligned with uncertainty and more aligned with danger \rev{perhaps due to its design which encodes low confidence with darker edges and may be misinterpreted as more solid or dire. Regardless of cause, our findings suggest} that overlaid blur affords different semantics than blur glyphs, which may contribute to its more cautious decision pattern.

\section{Discussion}
\label{sec:discussion}
The three experiments we present in this paper are the first to investigate methods for presenting time-series forecasts' quantitative and qualitative uncertainty together. %
Across these experiments, we observed a consistent pattern, where low forecaster confidence often made participants more likely to take preventative action when a forecast was just below the decision threshold, and less willing to commit to such preventative measures when the forecast was just above the threshold. This pattern suggests that low confidence made participants less certain about the action the forecast warranted. We further found that this effect held when forecaster confidence was expressed through text, blur glyphs, color hue glyphs, color hue overlays, and transparency overlays. Together, these results suggest that \rev{ textual confidence labels and their visual alternatives can change decision making in similar ways.}

Although we observed consistent response patterns across many encodings, intuitiveness ratings point to potentially different mechanisms underlying these effects. For example, participants ranked our color hues as aligning more with danger than uncertainty, suggesting that color-hue-based decisions could be shaped by perceived hazard rather than forecaster confidence alone. In contrast, text, blur glyphs, and transparency overlays aligned less naturally with interpretations of danger. These results suggest that qualitative confidence encodings may influence decisions through different semantic pathways, even when they lead to  similar behavioral effects. Such a pattern motivates future research aimed at disentangling not only what decisions participants make in these contexts, but \textit{how} those decisions are formed.

\subsection{Design Recommendations}

For two-dimensional forecast designers, our experimental evidence contributes to the following recommendations:

\begin{enumerate}
    \item \textbf{Low expert confidence in a forecast leads to less certain behavior near decision boundaries.} If a designer is visualizing a forecast with low expert confidence, presenting this qualitative information alongside the forecast can temper readers' decisions.

    \item \textbf{If a designer wants qualitative confidence levels to hold consistent weight in readers' decision-making,} regardless of forecast position, \textbf{they can express confidence through text captions, color hue glyphs or overlays, blur glyphs, or transparency overlays.} Expressing qualitative confidence with sets of transparent icons or with dark ``fog'' around the outside of a forecast can lead to larger changes in behavior across low, medium, and high levels of confidence below an optimal decision boundary, but smaller changes in behavior above the boundary.

    \item \textbf{Both red-yellow-green color hue encodings consistently led to high ratings of high-medium-low danger and uncertainty.} Designers with forecast contexts that include \rev{dangerous outcomes could use this color option to emphasize danger.}

    \item \textbf{The effects of medium qualitative confidence were not consistently different from those generated by high qualitative confidence.} Including a ``medium'' level may increase forecast sets' visual complexity without clear added benefit. %
\end{enumerate}

\rev{Forecasters can use text to convey qualitative confidence, but if they desire a visual solution, the recommendations above provide guidance on the effects of possible designs.}

\subsection{Limitations \& Future Work}
\label{sec:limitations}

Our work motivates further research of the communication of multiple forms of forecast uncertainty simultaneously. Future research can benefit the visualization community by examining alternative visualization techniques, data types, audiences, and decision contexts.
\rev{We used a non-expert population to %
ensure large sample sizes and maintain high statistical power.
However, %
even though the participants' decisions were linked to monetary incentives, our experiment cannot replicate the high stakes of a real-world PSPS decision. Prior work using a similar PSPS task found that different visual representations of uncertainty had similar effects on both experienced grid operators and on non-experts, but the grid operators had a much lower tolerance for risk overall~\cite{matzen2025preprint}. }\rev{Future research should assess whether experts respond differently to visual cues and whether the risk level leads to different decision patterns when qualitative uncertainty is present.}

Additionally, further research into methods of presenting qualitative confidence, including alternate channels and even new visualization designs, is warranted. %
When applying blur directly to forecast CIs in Experiment 3, we struggled to create a design that logically communicated the concept of uncertainty without fundamentally changing the CIs' shape or position. \rev{This difficulty is reflected in overlaid blur's intuitiveness ratings and could impact the effects we report.} %
We describe our challenge with this design process in \cref{sec:stimuli-exp3}, and encourage future research to explore new methods for uncertainty without interfering with positional encoding, such that we can more successfully represent qualitative confidence in time-series forecasting.

\rev{Participants also reported the intuitiveness of designs' semantic alignment after using the designs with a semantic legend. Although our results align with previous findings~\cite{maceachren2012visual}, the intuitiveness rankings we collected could be skewed by learning effects from earlier tasks.}

\rev{Finally, these experiments focused on communicating uncertainty in the context of making decisions about a dangerous risk. In this context, the alignment between the encodings and the concept of danger was important. Future research will be needed to assess how these types of encodings influence decisions in situations that do not involve danger, where different semantic associations may be more important.}

\section{Conclusion}
This work underscores an important point for both forecasting practice and the visualization community: not all decision-relevant uncertainty can be reduced to probabilities, intervals, or other numeric depictions. Forecasters often hold justified, difficult-to-quantify judgments about the quality of evidence, adequacy of models, and the reliability of their forecasts in context. Our results show that presenting these qualitative judgments alongside quantitative variability can meaningfully shape \rev{non-experts}’ decisions, and that qualitative confidence can often be conveyed visually, as well as through text. For visualization research, our work expands the study of time-series uncertainty communication beyond quantified uncertainty. For forecasting practice, it suggests \rev{the potential impact of qualitative forecast confidence on decision making.}

\section*{AI Use Disclosure}
We used Claude Sonnet 4.6 from Anthropic to assist with R and LaTeX programming and debugging. We reviewed all code line by line, and checked programming logic and output to ensure quality and reliability. Code is available in supplemental materials for peer review, as well.%

\section*{Supplemental Materials}
\label{sec:supplemental_materials}
All supplemental materials are available on OSF at \href{https://osf.io/7ya2c/overview}{https://osf.io/7ya2c/overview} under a {\ccby{} CC BY 4.0 license}.
In particular, they include (1) survey materials, (2) stimuli, (3) survey data, (4) data analyses, and (5) appendix tables.
Our preregistrations are available at \href{https://osf.io/8fcnb/overview?}{https://osf.io/8fcnb/overview} for Experiment 1, and\\ \href{https://osf.io/8cb7n/overview}{https://osf.io/8cb7n/overview} for Experiments 2 and 3.

\acknowledgments{
  This work was supported in part by a grant from the National Science Foundation (\#2428149) and in part by a grant from the Department of Energy Office of Electricity. Sandia National Laboratories is a multimission laboratory managed and operated by National Technology \& Engineering Solutions of Sandia, LLC, a wholly owned subsidiary of Honeywell International Inc., for the U.S. Department of Energy’s National Nuclear Security Administration under contract DE-NA0003525. This paper describes objective technical results and analysis. Any subjective views or opinions that might be expressed in the paper do not necessarily represent the views of the U.S. Department of Energy or the United States Government.
}

\bibliographystyle{abbrv-doi-hyperref}
\bibliography{template}

\clearpage
\onecolumn
\appendix %
\section{Appendix}
\vspace*{-\baselineskip}
\crefalias{section}{appendix} %
\definecolor{highlight}{HTML}{fff9c8}
\begin{table}[H]
    \centering
    \caption{\cref{eq:exp1} effect estimates. l\_CrI and u\_CrI denote lower and upper bound of the 95\% credible intervals. All values are in log-odds, and effects with 95\% credible intervals that exclude zero are highlighted.}
    \label{tab:effects}
    \begin{tabular}{llccc}
        \hline
        \textbf{Qual.~Confidence} & \textbf{\% Crossing} & \textbf{Mean Est.} & \textbf{l\_CrI} & \textbf{u\_CrI} \\
        \hline
        baseline - low  & 0p &  -0.101 & -0.492 &  0.283 \\
        high - low      & 0p &  -0.060 & -0.446 &  0.316 \\
        baseline - high & 0p &  -0.041 & -0.430 &  0.343 \\
        \hline
        baseline - low  & 5p &  -0.235 & -0.589 &  0.119 \\
        high - low      & 5p &  -0.214 & -0.574 &  0.145 \\
        baseline - high & 5p &  -0.021 & -0.384 &  0.353 \\
        \hline
\rowcolor{highlight} baseline - low  & 25p &  -0.610 & -0.871 & -0.350 \\
\rowcolor{highlight} high - low      & 25p &  -0.729 & -0.993 & -0.469 \\
        baseline - high & 25p &   0.118 & -0.160 &  0.390 \\
        \hline
       \rowcolor{highlight} baseline - low  & 50p &   0.298 &  0.035 &  0.556 \\
        \rowcolor{highlight} high - low      & 50p &   0.261 &  0.000 &  0.527 \\
        baseline - high & 50p &   0.037 & -0.226 &  0.303 \\
        \hline
       \rowcolor{highlight} baseline - low  & 75p &   0.537 &  0.249 &  0.849 \\
       \rowcolor{highlight} high - low      & 75p &   0.678 &  0.372 &  0.967 \\
        baseline - high & 75p &  -0.141 & -0.451 &  0.174 \\
        \hline
        baseline - low  & 95p &   0.127 & -0.210 &  0.462 \\
       \rowcolor{highlight} high - low      & 95p &   0.662 &  0.296 &  1.033 \\
      \rowcolor{highlight}  baseline - high & 95p &  -0.535 & -0.908 & -0.159 \\
        \hline
        low - baseline & 5p - 0p   &  0.134 & -0.394 &  0.658 \\
        \rowcolor{highlight} low - baseline & 25p - 0p  &  0.510 &  0.046 &  0.972 \\
        low - baseline & 50p - 0p  & -0.399 & -0.865 &  0.056 \\
        \rowcolor{highlight} low - baseline & 75p - 0p  & -0.637 & -1.118 & -0.176 \\
        low - baseline & 95p - 0p  & -0.227 & -0.732 &  0.286 \\
        low - baseline & 25p - 5p  &  0.375 & -0.057 &  0.815 \\
        \rowcolor{highlight} low - baseline & 50p - 5p  & -0.533 & -0.949 & -0.096 \\
       \rowcolor{highlight} low - baseline & 75p - 5p  & -0.772 & -1.232 & -0.297 \\
        low - baseline & 95p - 5p  & -0.362 & -0.859 &  0.128 \\
       \rowcolor{highlight} low - baseline & 50p - 25p & -0.908 & -1.277 & -0.542 \\
       \rowcolor{highlight} low - baseline & 75p - 25p & -1.147 & -1.536 & -0.751 \\
      \rowcolor{highlight}  low - baseline & 95p - 25p & -0.737 & -1.161 & -0.316 \\
        low - baseline & 75p - 50p & -0.239 & -0.634 &  0.158 \\
        low - baseline & 95p - 50p &  0.171 & -0.250 &  0.591 \\
        low - baseline & 95p - 75p &  0.410 & -0.034 &  0.862 \\
        \hline
        high - baseline & 5p - 0p   & -0.019 & -0.558 &  0.508 \\
        high - baseline & 25p - 0p  & -0.159 & -0.642 &  0.308 \\
        high - baseline & 50p - 0p  & -0.078 & -0.543 &  0.396 \\
        high - baseline & 75p - 0p  &  0.100 & -0.392 &  0.601 \\
        high - baseline & 95p - 0p  &  0.495 & -0.039 &  1.024 \\
        high - baseline & 25p - 5p  & -0.140 & -0.599 &  0.319 \\
        high - baseline & 50p - 5p  & -0.059 & -0.501 &  0.400 \\
        high - baseline & 75p - 5p  &  0.119 & -0.365 &  0.608 \\
        high - baseline & 95p - 5p  &  0.514 & -0.012 &  1.052 \\
        high - baseline & 50p - 25p &  0.081 & -0.314 &  0.475 \\
        high - baseline & 75p - 25p &  0.259 & -0.163 &  0.674 \\
       \rowcolor{highlight} high - baseline & 95p - 25p &  0.654 &  0.197 &  1.109 \\
        high - baseline & 75p - 50p &  0.178 & -0.235 &  0.591 \\
      \rowcolor{highlight}  high - baseline & 95p - 50p &  0.573 &  0.123 &  1.027 \\
        high - baseline & 95p - 75p &  0.395 & -0.089 &  0.878 \\
        \hline
        high - low & 5p - 0p   & -0.154 & -0.676 &  0.348 \\
       \rowcolor{highlight} high - low & 25p - 0p  & -0.669 & -1.137 & -0.228 \\
        high - low & 50p - 0p  &  0.321 & -0.141 &  0.785 \\
       \rowcolor{highlight} high - low & 75p - 0p  &  0.737 &  0.275 &  1.208 \\
       \rowcolor{highlight} high - low & 95p - 0p  &  0.722 &  0.204 &  1.256 \\
      \rowcolor{highlight}  high - low & 25p - 5p  & -0.515 & -0.955 & -0.062 \\
       \rowcolor{highlight} high - low & 50p - 5p  &  0.475 &  0.031 &  0.936 \\
       \rowcolor{highlight} high - low & 75p - 5p  &  0.891 &  0.421 &  1.370 \\
      \rowcolor{highlight}  high - low & 95p - 5p  &  0.876 &  0.361 &  1.393 \\
       \rowcolor{highlight} high - low & 50p - 25p &  0.990 &  0.621 &  1.365 \\
      \rowcolor{highlight}  high - low & 75p - 25p &  1.406 &  1.015 &  1.802 \\
       \rowcolor{highlight} high - low & 95p - 25p &  1.391 &  0.939 &  1.834 \\
      \rowcolor{highlight}  high - low & 75p - 50p &  0.417 &  0.007 &  0.803 \\
        high - low & 95p - 50p &  0.401 & -0.058 &  0.861 \\
        high - low & 95p - 75p & -0.015 & -0.472 &  0.453 \\
        \hline
    \end{tabular}
    \label{tab:exp1-ests}
\end{table}
\clearpage

\begin{center}
\begin{longtable}{lllccc}
    \caption{Interaction effect estimates from \cref{eq:exp2} applied to Experiment 2 data. l\_CrI and u\_CrI denote lower and upper bound of the 95\% credible intervals. All values are in log-odds, and effects with 95\% credible intervals that exclude zero are highlighted. \textit{g\_\{condition\}} denotes glyph versions of encoding conditions.}
    \label{tab:effects2} \\
    \hline
    \textbf{Qual.~Confidence} & \textbf{Confidence Encoding} & \textbf{\% Crossing} & \textbf{Mean Est.} & \textbf{l\_CrI} & \textbf{u\_CrI} \\
    \hline
    \endfirsthead

    \multicolumn{6}{c}{\tablename\ \thetable{} -- continued from previous page} \\
    \hline
   \textbf{Qual.~Confidence} & \textbf{Confidence Encoding} & \textbf{\% Crossing} & \textbf{Mean Est.} & \textbf{l\_CrI} & \textbf{u\_CrI} \\
    \hline
    \endhead

    \hline
    \multicolumn{6}{r}{Continued on next page} \\
    \endfoot

    \hline
    \endlastfoot

    med - low & g\_color - text      & 5p  &  0.031 & -0.406 &  0.464 \\
    med - low & g\_blur - text       & 5p  & -0.086 & -0.507 &  0.347 \\
    med - low & g\_transp - text     & 5p  & -0.181 & -0.589 &  0.234 \\
    med - low & g\_blur - g\_color   & 5p  & -0.116 & -0.546 &  0.304 \\
    med - low & g\_transp - g\_color & 5p  & -0.212 & -0.642 &  0.184 \\
    med - low & g\_transp - g\_blur  & 5p  & -0.096 & -0.502 &  0.318 \\
    med - low & g\_color - text      & 25p &  0.031 & -0.406 &  0.464 \\
    med - low & g\_blur - text       & 25p & -0.086 & -0.507 &  0.347 \\
    med - low & g\_transp - text     & 25p & -0.181 & -0.589 &  0.234 \\
    med - low & g\_blur - g\_color   & 25p & -0.116 & -0.546 &  0.304 \\
    med - low & g\_transp - g\_color & 25p & -0.212 & -0.642 &  0.184 \\
    med - low & g\_transp - g\_blur  & 25p & -0.096 & -0.502 &  0.318 \\
    med - low & g\_color - text      & 50p &  0.031 & -0.406 &  0.464 \\
    med - low & g\_blur - text       & 50p & -0.086 & -0.507 &  0.347 \\
    med - low & g\_transp - text     & 50p & -0.181 & -0.589 &  0.234 \\
    med - low & g\_blur - g\_color   & 50p & -0.116 & -0.546 &  0.304 \\
    med - low & g\_transp - g\_color & 50p & -0.212 & -0.642 &  0.184 \\
    med - low & g\_transp - g\_blur  & 50p & -0.096 & -0.502 &  0.318 \\
    med - low & g\_color - text      & 75p &  0.031 & -0.406 &  0.464 \\
    med - low & g\_blur - text       & 75p & -0.086 & -0.507 &  0.347 \\
    med - low & g\_transp - text     & 75p & -0.181 & -0.589 &  0.234 \\
    med - low & g\_blur - g\_color   & 75p & -0.116 & -0.546 &  0.304 \\
    med - low & g\_transp - g\_color & 75p & -0.212 & -0.642 &  0.184 \\
    med - low & g\_transp - g\_blur  & 75p & -0.096 & -0.502 &  0.318 \\
    \hline
    high - low & g\_color - text      & 5p  & -0.273 & -0.700 &  0.139 \\
    high - low & g\_blur - text       & 5p  & -0.277 & -0.702 &  0.149 \\
    \rowcolor{highlight}
    high - low & g\_transp - text     & 5p  & -0.936 & -1.356 & -0.515 \\
    high - low & g\_blur - g\_color   & 5p  & -0.004 & -0.436 &  0.415 \\
    \rowcolor{highlight}
    high - low & g\_transp - g\_color & 5p  & -0.663 & -1.076 & -0.239 \\
    \rowcolor{highlight}
    high - low & g\_transp - g\_blur  & 5p  & -0.659 & -1.068 & -0.250 \\
    high - low & g\_color - text      & 25p & -0.273 & -0.700 &  0.139 \\
    high - low & g\_blur - text       & 25p & -0.277 & -0.702 &  0.149 \\
    \rowcolor{highlight}
    high - low & g\_transp - text     & 25p & -0.936 & -1.356 & -0.515 \\
    high - low & g\_blur - g\_color   & 25p & -0.004 & -0.436 &  0.415 \\
    \rowcolor{highlight}
    high - low & g\_transp - g\_color & 25p & -0.663 & -1.076 & -0.239 \\
    \rowcolor{highlight}
    high - low & g\_transp - g\_blur  & 25p & -0.659 & -1.068 & -0.250 \\
    high - low & g\_color - text      & 50p & -0.273 & -0.700 &  0.139 \\
    high - low & g\_blur - text       & 50p & -0.277 & -0.702 &  0.149 \\
    \rowcolor{highlight}
    high - low & g\_transp - text     & 50p & -0.936 & -1.356 & -0.515 \\
    high - low & g\_blur - g\_color   & 50p & -0.004 & -0.436 &  0.415 \\
    \rowcolor{highlight}
    high - low & g\_transp - g\_color & 50p & -0.663 & -1.076 & -0.239 \\
    \rowcolor{highlight}
    high - low & g\_transp - g\_blur  & 50p & -0.659 & -1.068 & -0.250 \\
    high - low & g\_color - text      & 75p & -0.273 & -0.700 &  0.139 \\
    high - low & g\_blur - text       & 75p & -0.277 & -0.702 &  0.149 \\
    \rowcolor{highlight}
    high - low & g\_transp - text     & 75p & -0.936 & -1.356 & -0.515 \\
    high - low & g\_blur - g\_color   & 75p & -0.004 & -0.436 &  0.415 \\
    \rowcolor{highlight}
    high - low & g\_transp - g\_color & 75p & -0.663 & -1.076 & -0.239 \\
    \rowcolor{highlight}
    high - low & g\_transp - g\_blur  & 75p & -0.659 & -1.068 & -0.250 \\
    \hline
    high - med & g\_color - text      & 5p  & -0.304 & -0.742 &  0.153 \\
    high - med & g\_blur - text       & 5p  & -0.192 & -0.636 &  0.236 \\
    \rowcolor{highlight}
    high - med & g\_transp - text     & 5p  & -0.755 & -1.198 & -0.322 \\
    high - med & g\_blur - g\_color   & 5p  &  0.113 & -0.318 &  0.523 \\
    \rowcolor{highlight}
    high - med & g\_transp - g\_color & 5p  & -0.451 & -0.870 & -0.018 \\
    \rowcolor{highlight}
    high - med & g\_transp - g\_blur  & 5p  & -0.563 & -0.994 & -0.128 \\
    high - med & g\_color - text      & 25p & -0.304 & -0.742 &  0.153 \\
    high - med & g\_blur - text       & 25p & -0.192 & -0.636 &  0.236 \\
    \rowcolor{highlight}
    high - med & g\_transp - text     & 25p & -0.755 & -1.198 & -0.322 \\
    high - med & g\_blur - g\_color   & 25p &  0.113 & -0.318 &  0.523 \\
    \rowcolor{highlight}
    high - med & g\_transp - g\_color & 25p & -0.451 & -0.870 & -0.018 \\
    \rowcolor{highlight}
    high - med & g\_transp - g\_blur  & 25p & -0.563 & -0.994 & -0.128 \\
    high - med & g\_color - text      & 50p & -0.304 & -0.742 &  0.153 \\
    high - med & g\_blur - text       & 50p & -0.192 & -0.636 &  0.236 \\
    \rowcolor{highlight}
    high - med & g\_transp - text     & 50p & -0.755 & -1.198 & -0.322 \\
    high - med & g\_blur - g\_color   & 50p &  0.113 & -0.318 &  0.523 \\
    \rowcolor{highlight}
    high - med & g\_transp - g\_color & 50p & -0.451 & -0.870 & -0.018 \\
    \rowcolor{highlight}
    high - med & g\_transp - g\_blur  & 50p & -0.563 & -0.994 & -0.128 \\
    high - med & g\_color - text      & 75p & -0.304 & -0.742 &  0.153 \\
    high - med & g\_blur - text       & 75p & -0.192 & -0.636 &  0.236 \\
    \rowcolor{highlight}
    high - med & g\_transp - text     & 75p & -0.755 & -1.198 & -0.322 \\
    high - med & g\_blur - g\_color   & 75p &  0.113 & -0.318 &  0.523 \\
    \rowcolor{highlight}
    high - med & g\_transp - g\_color & 75p & -0.451 & -0.870 & -0.018 \\
    \rowcolor{highlight}
    high - med & g\_transp - g\_blur  & 75p & -0.563 & -0.994 & -0.128 \\
    \label{tab:exp2-ests}
\end{longtable}
\end{center}

\begin{center}
\begin{longtable}{lllccc}
    \caption{Interaction effect estimates from \cref{eq:exp2} applied to Experiment 3 data. l\_CrI and u\_CrI denote lower and upper bound of the 95\% credible intervals. All values are in log-odds, and effects with 95\% credible intervals that exclude zero are highlighted. \textit{ol\_\{condition\}} denotes overlaid versions of encoding conditions.}
    \label{tab:effects2} \\
    \hline
    \textbf{Qual.~Confidence} & \textbf{Confidence Encoding} & \textbf{\% Crossing} & \textbf{Mean Est.} & \textbf{l\_CrI} & \textbf{u\_CrI} \\
    \hline
    \endfirsthead

    \multicolumn{6}{c}{\tablename\ \thetable{} -- continued from previous page} \\
    \hline
   \textbf{Qual.~Confidence} & \textbf{Confidence Encoding} & \textbf{\% Crossing} & \textbf{Mean Est.} & \textbf{l\_CrI} & \textbf{u\_CrI} \\
    \hline
    \endhead

    \hline
    \multicolumn{6}{r}{Continued on next page} \\
    \endfoot

    \hline
    \endlastfoot

    med - low & ol\_color - text       & 5p  & -0.149 & -0.566 &  0.274 \\
    med - low & ol\_blur - text        & 5p  & -0.359 & -0.767 &  0.051 \\
    med - low & ol\_transp - text      & 5p  &  0.258 & -0.156 &  0.672 \\
    med - low & ol\_blur - ol\_color   & 5p  & -0.211 & -0.620 &  0.213 \\
    med - low & ol\_transp - ol\_color & 5p  &  0.406 & -0.017 &  0.830 \\
    \rowcolor{highlight}
    med - low & ol\_transp - ol\_blur  & 5p  &  0.617 &  0.195 &  1.037 \\
    med - low & ol\_color - text       & 25p & -0.149 & -0.566 &  0.274 \\
    med - low & ol\_blur - text        & 25p & -0.359 & -0.767 &  0.051 \\
    med - low & ol\_transp - text      & 25p &  0.258 & -0.156 &  0.672 \\
    med - low & ol\_blur - ol\_color   & 25p & -0.211 & -0.620 &  0.213 \\
    med - low & ol\_transp - ol\_color & 25p &  0.406 & -0.017 &  0.830 \\
    \rowcolor{highlight}
    med - low & ol\_transp - ol\_blur  & 25p &  0.617 &  0.195 &  1.037 \\
    med - low & ol\_color - text       & 50p & -0.149 & -0.566 &  0.274 \\
    med - low & ol\_blur - text        & 50p & -0.359 & -0.767 &  0.051 \\
    med - low & ol\_transp - text      & 50p &  0.258 & -0.156 &  0.672 \\
    med - low & ol\_blur - ol\_color   & 50p & -0.211 & -0.620 &  0.213 \\
    med - low & ol\_transp - ol\_color & 50p &  0.406 & -0.017 &  0.830 \\
    \rowcolor{highlight}
    med - low & ol\_transp - ol\_blur  & 50p &  0.617 &  0.195 &  1.037 \\
    med - low & ol\_color - text       & 75p & -0.149 & -0.566 &  0.274 \\
    med - low & ol\_blur - text        & 75p & -0.359 & -0.767 &  0.051 \\
    med - low & ol\_transp - text      & 75p &  0.258 & -0.156 &  0.672 \\
    med - low & ol\_blur - ol\_color   & 75p & -0.211 & -0.620 &  0.213 \\
    med - low & ol\_transp - ol\_color & 75p &  0.406 & -0.017 &  0.830 \\
    \rowcolor{highlight}
    med - low & ol\_transp - ol\_blur  & 75p &  0.617 &  0.195 &  1.037 \\
    \hline
    high - low & ol\_color - text       & 5p  & -0.261 & -0.705 &  0.157 \\
    \rowcolor{highlight}
    high - low & ol\_blur - text        & 5p  & -0.635 & -1.064 & -0.216 \\
    high - low & ol\_transp - text      & 5p  &  0.155 & -0.275 &  0.579 \\
    high - low & ol\_blur - ol\_color   & 5p  & -0.374 & -0.802 &  0.048 \\
    high - low & ol\_transp - ol\_color & 5p  &  0.415 & -0.022 &  0.851 \\
    \rowcolor{highlight}
    high - low & ol\_transp - ol\_blur  & 5p  &  0.789 &  0.365 &  1.224 \\
    high - low & ol\_color - text       & 25p & -0.261 & -0.705 &  0.157 \\
    \rowcolor{highlight}
    high - low & ol\_blur - text        & 25p & -0.635 & -1.064 & -0.216 \\
    high - low & ol\_transp - text      & 25p &  0.155 & -0.275 &  0.579 \\
    high - low & ol\_blur - ol\_color   & 25p & -0.374 & -0.802 &  0.048 \\
    high - low & ol\_transp - ol\_color & 25p &  0.415 & -0.022 &  0.851 \\
    \rowcolor{highlight}
    high - low & ol\_transp - ol\_blur  & 25p &  0.789 &  0.365 &  1.224 \\
    high - low & ol\_color - text       & 50p & -0.261 & -0.705 &  0.157 \\
    \rowcolor{highlight}
    high - low & ol\_blur - text        & 50p & -0.635 & -1.064 & -0.216 \\
    high - low & ol\_transp - text      & 50p &  0.155 & -0.275 &  0.579 \\
    high - low & ol\_blur - ol\_color   & 50p & -0.374 & -0.802 &  0.048 \\
    high - low & ol\_transp - ol\_color & 50p &  0.415 & -0.022 &  0.851 \\
    \rowcolor{highlight}
    high - low & ol\_transp - ol\_blur  & 50p &  0.789 &  0.365 &  1.224 \\
    high - low & ol\_color - text       & 75p & -0.261 & -0.705 &  0.157 \\
    \rowcolor{highlight}
    high - low & ol\_blur - text        & 75p & -0.635 & -1.064 & -0.216 \\
    high - low & ol\_transp - text      & 75p &  0.155 & -0.275 &  0.579 \\
    high - low & ol\_blur - ol\_color   & 75p & -0.374 & -0.802 &  0.048 \\
    high - low & ol\_transp - ol\_color & 75p &  0.415 & -0.022 &  0.851 \\
    \rowcolor{highlight}
    high - low & ol\_transp - ol\_blur  & 75p &  0.789 &  0.365 &  1.224 \\
    \hline
    high - med & ol\_color - text       & 5p  & -0.112 & -0.555 &  0.321 \\
    high - med & ol\_blur - text        & 5p  & -0.275 & -0.688 &  0.155 \\
    high - med & ol\_transp - text      & 5p  & -0.103 & -0.523 &  0.321 \\
    high - med & ol\_blur - ol\_color   & 5p  & -0.163 & -0.603 &  0.271 \\
    high - med & ol\_transp - ol\_color & 5p  &  0.009 & -0.431 &  0.437 \\
    high - med & ol\_transp - ol\_blur  & 5p  &  0.172 & -0.265 &  0.615 \\
    high - med & ol\_color - text       & 25p & -0.112 & -0.555 &  0.321 \\
    high - med & ol\_blur - text        & 25p & -0.275 & -0.688 &  0.155 \\
    high - med & ol\_transp - text      & 25p & -0.103 & -0.523 &  0.321 \\
    high - med & ol\_blur - ol\_color   & 25p & -0.163 & -0.603 &  0.271 \\
    high - med & ol\_transp - ol\_color & 25p &  0.009 & -0.431 &  0.437 \\
    high - med & ol\_transp - ol\_blur  & 25p &  0.172 & -0.265 &  0.615 \\
    high - med & ol\_color - text       & 50p & -0.112 & -0.555 &  0.321 \\
    high - med & ol\_blur - text        & 50p & -0.275 & -0.688 &  0.155 \\
    high - med & ol\_transp - text      & 50p & -0.103 & -0.523 &  0.321 \\
    high - med & ol\_blur - ol\_color   & 50p & -0.163 & -0.603 &  0.271 \\
    high - med & ol\_transp - ol\_color & 50p &  0.009 & -0.431 &  0.437 \\
    high - med & ol\_transp - ol\_blur  & 50p &  0.172 & -0.265 &  0.615 \\
    high - med & ol\_color - text       & 75p & -0.112 & -0.555 &  0.321 \\
    high - med & ol\_blur - text        & 75p & -0.275 & -0.688 &  0.155 \\
    high - med & ol\_transp - text      & 75p & -0.103 & -0.523 &  0.321 \\
    high - med & ol\_blur - ol\_color   & 75p & -0.163 & -0.603 &  0.271 \\
    high - med & ol\_transp - ol\_color & 75p &  0.009 & -0.431 &  0.437 \\
    high - med & ol\_transp - ol\_blur  & 75p &  0.172 & -0.265 &  0.615 \\
    \label{tab:exp3-ests}
\end{longtable}
\end{center}

\begin{center}
\begin{longtable}{llccc}
    \caption{Interaction effect estimates from \cref{eq:intuit} applied to Experiment 2 and 3 intuitiveness rating data. l\_CrI and u\_CrI denote lower and upper bound of the 95\% credible intervals. All values are in log-odds, and effects with 95\% credible intervals that exclude zero are highlighted. \textit{g\_\{condition\}} and  \textit{ol\_\{condition\}} denotes glyph and overlaid versions of encoding conditions.}\\
    \hline
    \textbf{Confidence Encoding} & \textbf{Rating Question} & \textbf{Mean Est.} & \textbf{l\_CrI} & \textbf{u\_CrI} \\
    \hline
    \endfirsthead
    \multicolumn{5}{c}{\tablename\ \thetable{} -- continued from previous page} \\
    \hline
    \textbf{Confidence Encoding} & \textbf{Rating Question} & \textbf{Mean Est.} & \textbf{l\_CrI} & \textbf{u\_CrI} \\
    \hline
    \endhead
    \hline
    \multicolumn{5}{r}{Continued on next page} \\
    \endfoot
    \hline
    \endlastfoot

    \rowcolor{highlight}
    g\_blur - text & certain.aligned  & -0.767 & -1.251 & -0.293 \\
    g\_blur - text & certain.opposite &  0.367 & -0.274 &  0.993 \\
    \rowcolor{highlight}
    g\_blur - text & danger.aligned   & -1.603 & -2.255 & -0.933 \\
    \rowcolor{highlight}
    g\_blur - text & danger.opposite  &  1.193 &  0.533 &  1.840 \\
    \hline
    g\_color - text & certain.aligned  & -0.344 & -0.765 &  0.076 \\
    g\_color - text & certain.opposite &  0.570 & -0.085 &  1.223 \\
    \rowcolor{highlight}
    g\_color - text & danger.aligned   &  1.680 &  1.126 &  2.237 \\
    \rowcolor{highlight}
    g\_color - text & danger.opposite  & -1.344 & -1.999 & -0.665 \\
    \hline
    \rowcolor{highlight}
    g\_transp - text & certain.aligned  & -0.539 & -0.986 & -0.099 \\
    g\_transp - text & certain.opposite &  0.324 & -0.315 &  0.960 \\
    \rowcolor{highlight}
    g\_transp - text & danger.aligned   & -2.005 & -2.633 & -1.353 \\
    \rowcolor{highlight}
    g\_transp - text & danger.opposite  &  1.872 &  1.260 &  2.463 \\
    \hline
    \rowcolor{highlight}
    ol\_blur - text & certain.aligned  & -1.480 & -2.004 & -0.962 \\
    \rowcolor{highlight}
    ol\_blur - text & certain.opposite &  1.521 &  0.861 &  2.151 \\
    ol\_blur - text & danger.aligned   &  0.454 & -0.181 &  1.093 \\
    ol\_blur - text & danger.opposite  & -0.624 & -1.280 &  0.054 \\
    \hline
    ol\_color - text & certain.aligned  & -0.107 & -0.508 &  0.292 \\
    ol\_color - text & certain.opposite & -0.039 & -0.657 &  0.583 \\
    \rowcolor{highlight}
    ol\_color - text & danger.aligned   &  1.179 &  0.563 &  1.787 \\
    \rowcolor{highlight}
    ol\_color - text & danger.opposite  & -1.335 & -2.009 & -0.644 \\
    \hline
    \rowcolor{highlight}
    ol\_transp - text & certain.aligned  & -0.491 & -0.933 & -0.058 \\
    ol\_transp - text & certain.opposite &  0.076 & -0.551 &  0.705 \\
    \rowcolor{highlight}
    ol\_transp - text & danger.aligned   & -1.617 & -2.268 & -0.939 \\
    \rowcolor{highlight}
    ol\_transp - text & danger.opposite  &  0.863 &  0.172 &  1.544 \\
    \hline
    g\_blur - g\_color & certain.aligned  & -0.423 & -0.928 &  0.067 \\
    g\_blur - g\_color & certain.opposite & -0.203 & -0.873 &  0.467 \\
    \rowcolor{highlight}
    g\_blur - g\_color & danger.aligned   & -3.283 & -3.796 & -2.744 \\
    \rowcolor{highlight}
    g\_blur - g\_color & danger.opposite  &  2.537 &  1.889 &  3.139 \\
    \hline
    g\_transp - g\_color & certain.aligned  & -0.195 & -0.670 &  0.268 \\
    g\_transp - g\_color & certain.opposite & -0.246 & -0.908 &  0.411 \\
    \rowcolor{highlight}
    g\_transp - g\_color & danger.aligned   & -3.685 & -4.160 & -3.177 \\
    \rowcolor{highlight}
    g\_transp - g\_color & danger.opposite  &  3.216 &  2.639 &  3.752 \\
    \hline
    \rowcolor{highlight}
    ol\_blur - g\_color & certain.aligned  & -1.136 & -1.676 & -0.589 \\
    \rowcolor{highlight}
    ol\_blur - g\_color & certain.opposite &  0.951 &  0.267 &  1.609 \\
    \rowcolor{highlight}
    ol\_blur - g\_color & danger.aligned   & -1.226 & -1.731 & -0.724 \\
    \rowcolor{highlight}
    ol\_blur - g\_color & danger.opposite  &  0.720 &  0.074 &  1.345 \\
    \hline
    ol\_color - g\_color & certain.aligned  &  0.237 & -0.191 &  0.669 \\
    ol\_color - g\_color & certain.opposite & -0.609 & -1.259 &  0.049 \\
    \rowcolor{highlight}
    ol\_color - g\_color & danger.aligned   & -0.501 & -0.980 & -0.039 \\
    ol\_color - g\_color & danger.opposite  &  0.009 & -0.642 &  0.655 \\
    \hline
    ol\_transp - g\_color & certain.aligned  & -0.147 & -0.626 &  0.321 \\
    ol\_transp - g\_color & certain.opposite & -0.494 & -1.146 &  0.171 \\
    \rowcolor{highlight}
    ol\_transp - g\_color & danger.aligned   & -3.297 & -3.812 & -2.748 \\
    \rowcolor{highlight}
    ol\_transp - g\_color & danger.opposite  &  2.207 &  1.534 &  2.834 \\
    \hline
    text - g\_color & certain.aligned  &  0.344 & -0.076 &  0.765 \\
    text - g\_color & certain.opposite & -0.570 & -1.223 &  0.085 \\
    \rowcolor{highlight}
    text - g\_color & danger.aligned   & -1.680 & -2.237 & -1.126 \\
    \rowcolor{highlight}
    text - g\_color & danger.opposite  &  1.344 &  0.665 &  1.999 \\
    \hline
    \rowcolor{highlight}
    g\_blur - ol\_color & certain.aligned  & -0.660 & -1.144 & -0.180 \\
    g\_blur - ol\_color & certain.opposite &  0.407 & -0.226 &  1.034 \\
    \rowcolor{highlight}
    g\_blur - ol\_color & danger.aligned   & -2.782 & -3.351 & -2.187 \\
    \rowcolor{highlight}
    g\_blur - ol\_color & danger.opposite  &  2.528 &  1.868 &  3.152 \\
    \hline
    g\_color - ol\_color & certain.aligned  & -0.237 & -0.669 &  0.191 \\
    g\_color - ol\_color & certain.opposite &  0.609 & -0.049 &  1.259 \\
    \rowcolor{highlight}
    g\_color - ol\_color & danger.aligned   &  0.501 &  0.039 &  0.980 \\
    g\_color - ol\_color & danger.opposite  & -0.009 & -0.655 &  0.642 \\
    \hline
    g\_transp - ol\_color & certain.aligned  & -0.432 & -0.890 &  0.020 \\
    g\_transp - ol\_color & certain.opposite &  0.363 & -0.272 &  1.005 \\
    \rowcolor{highlight}
    g\_transp - ol\_color & danger.aligned   & -3.184 & -3.731 & -2.600 \\
    \rowcolor{highlight}
    g\_transp - ol\_color & danger.opposite  &  3.207 &  2.613 &  3.751 \\
    \hline
    \rowcolor{highlight}
    ol\_blur - ol\_color & certain.aligned  & -1.373 & -1.900 & -0.841 \\
    \rowcolor{highlight}
    ol\_blur - ol\_color & certain.opposite &  1.560 &  0.895 &  2.197 \\
    \rowcolor{highlight}
    ol\_blur - ol\_color & danger.aligned   & -0.725 & -1.294 & -0.152 \\
    \rowcolor{highlight}
    ol\_blur - ol\_color & danger.opposite  &  0.711 &  0.052 &  1.342 \\
    \hline
    ol\_transp - ol\_color & certain.aligned  & -0.384 & -0.847 &  0.070 \\
    ol\_transp - ol\_color & certain.opposite &  0.115 & -0.510 &  0.751 \\
    \rowcolor{highlight}
    ol\_transp - ol\_color & danger.aligned   & -2.796 & -3.368 & -2.187 \\
    \rowcolor{highlight}
    ol\_transp - ol\_color & danger.opposite  &  2.198 &  1.508 &  2.839 \\
    \hline
    text - ol\_color & certain.aligned  &  0.107 & -0.292 &  0.508 \\
    text - ol\_color & certain.opposite &  0.039 & -0.583 &  0.657 \\
    \rowcolor{highlight}
    text - ol\_color & danger.aligned   & -1.179 & -1.787 & -0.563 \\
    \rowcolor{highlight}
    text - ol\_color & danger.opposite  &  1.335 &  0.644 &  2.009 \\
    \hline
    g\_color - g\_blur & certain.aligned  &  0.423 & -0.067 &  0.928 \\
    g\_color - g\_blur & certain.opposite &  0.203 & -0.467 &  0.873 \\
    \rowcolor{highlight}
    g\_color - g\_blur & danger.aligned   &  3.283 &  2.744 &  3.796 \\
    \rowcolor{highlight}
    g\_color - g\_blur & danger.opposite  & -2.537 & -3.139 & -1.889 \\
    \hline
    g\_transp - g\_blur & certain.aligned  &  0.228 & -0.289 &  0.758 \\
    g\_transp - g\_blur & certain.opposite & -0.043 & -0.698 &  0.601 \\
    g\_transp - g\_blur & danger.aligned   & -0.402 & -1.015 &  0.209 \\
    \rowcolor{highlight}
    g\_transp - g\_blur & danger.opposite  &  0.679 &  0.130 &  1.224 \\
    \hline
    \rowcolor{highlight}
    ol\_blur - g\_blur & certain.aligned  & -0.713 & -1.298 & -0.115 \\
    \rowcolor{highlight}
    ol\_blur - g\_blur & certain.opposite &  1.153 &  0.484 &  1.808 \\
    \rowcolor{highlight}
    ol\_blur - g\_blur & danger.aligned   &  2.057 &  1.430 &  2.661 \\
    \rowcolor{highlight}
    ol\_blur - g\_blur & danger.opposite  & -1.818 & -2.418 & -1.185 \\
    \hline
    \rowcolor{highlight}
    ol\_color - g\_blur & certain.aligned  &  0.660 &  0.180 &  1.144 \\
    ol\_color - g\_blur & certain.opposite & -0.407 & -1.034 &  0.226 \\
    \rowcolor{highlight}
    ol\_color - g\_blur & danger.aligned   &  2.782 &  2.187 &  3.351 \\
    \rowcolor{highlight}
    ol\_color - g\_blur & danger.opposite  & -2.528 & -3.152 & -1.868 \\
    \hline
    ol\_transp - g\_blur & certain.aligned  &  0.276 & -0.245 &  0.805 \\
    ol\_transp - g\_blur & certain.opposite & -0.292 & -0.929 &  0.348 \\
    ol\_transp - g\_blur & danger.aligned   & -0.014 & -0.649 &  0.628 \\
    ol\_transp - g\_blur & danger.opposite  & -0.330 & -0.961 &  0.304 \\
    \hline
    \rowcolor{highlight}
    text - g\_blur & certain.aligned  &  0.767 &  0.293 &  1.251 \\
    text - g\_blur & certain.opposite & -0.367 & -0.993 &  0.274 \\
    \rowcolor{highlight}
    text - g\_blur & danger.aligned   &  1.603 &  0.933 &  2.255 \\
    \rowcolor{highlight}
    text - g\_blur & danger.opposite  & -1.193 & -1.840 & -0.533 \\
    \hline
    \rowcolor{highlight}
    g\_blur - ol\_blur & certain.aligned  &  0.713 &  0.115 &  1.298 \\
    \rowcolor{highlight}
    g\_blur - ol\_blur & certain.opposite & -1.153 & -1.808 & -0.484 \\
    \rowcolor{highlight}
    g\_blur - ol\_blur & danger.aligned   & -2.057 & -2.661 & -1.430 \\
    \rowcolor{highlight}
    g\_blur - ol\_blur & danger.opposite  &  1.818 &  1.185 &  2.418 \\
    \hline
    \rowcolor{highlight}
    g\_color - ol\_blur & certain.aligned  &  1.136 &  0.589 &  1.676 \\
    \rowcolor{highlight}
    g\_color - ol\_blur & certain.opposite & -0.951 & -1.609 & -0.267 \\
    \rowcolor{highlight}
    g\_color - ol\_blur & danger.aligned   &  1.226 &  0.724 &  1.731 \\
    \rowcolor{highlight}
    g\_color - ol\_blur & danger.opposite  & -0.720 & -1.345 & -0.074 \\
    \hline
    \rowcolor{highlight}
    g\_transp - ol\_blur & certain.aligned  &  0.941 &  0.381 &  1.504 \\
    \rowcolor{highlight}
    g\_transp - ol\_blur & certain.opposite & -1.197 & -1.846 & -0.521 \\
    \rowcolor{highlight}
    g\_transp - ol\_blur & danger.aligned   & -2.459 & -3.038 & -1.839 \\
    \rowcolor{highlight}
    g\_transp - ol\_blur & danger.opposite  &  2.496 &  1.922 &  3.045 \\
    \hline
    \rowcolor{highlight}
    ol\_color - ol\_blur & certain.aligned  &  1.373 &  0.841 &  1.900 \\
    \rowcolor{highlight}
    ol\_color - ol\_blur & certain.opposite & -1.560 & -2.197 & -0.895 \\
    \rowcolor{highlight}
    ol\_color - ol\_blur & danger.aligned   &  0.725 &  0.152 &  1.294 \\
    \rowcolor{highlight}
    ol\_color - ol\_blur & danger.opposite  & -0.711 & -1.342 & -0.052 \\
    \hline
    \rowcolor{highlight}
    ol\_transp - ol\_blur & certain.aligned  &  0.989 &  0.424 &  1.547 \\
    \rowcolor{highlight}
    ol\_transp - ol\_blur & certain.opposite & -1.445 & -2.085 & -0.774 \\
    \rowcolor{highlight}
    ol\_transp - ol\_blur & danger.aligned   & -2.071 & -2.671 & -1.439 \\
    \rowcolor{highlight}
    ol\_transp - ol\_blur & danger.opposite  &  1.487 &  0.821 &  2.120 \\
    \hline
    \rowcolor{highlight}
    text - ol\_blur & certain.aligned  &  1.480 &  0.962 &  2.004 \\
    \rowcolor{highlight}
    text - ol\_blur & certain.opposite & -1.521 & -2.151 & -0.861 \\
    text - ol\_blur & danger.aligned   & -0.454 & -1.093 &  0.181 \\
    text - ol\_blur & danger.opposite  &  0.624 & -0.054 &  1.280 \\
    \hline
    g\_blur - g\_transp & certain.aligned  & -0.228 & -0.758 &  0.289 \\
    g\_blur - g\_transp & certain.opposite &  0.043 & -0.601 &  0.698 \\
    g\_blur - g\_transp & danger.aligned   &  0.402 & -0.209 &  1.015 \\
    \rowcolor{highlight}
    g\_blur - g\_transp & danger.opposite  & -0.679 & -1.224 & -0.130 \\
    \hline
    g\_color - g\_transp & certain.aligned  &  0.195 & -0.268 &  0.670 \\
    g\_color - g\_transp & certain.opposite &  0.246 & -0.411 &  0.908 \\
    \rowcolor{highlight}
    g\_color - g\_transp & danger.aligned   &  3.685 &  3.177 &  4.160 \\
    \rowcolor{highlight}
    g\_color - g\_transp & danger.opposite  & -3.216 & -3.752 & -2.639 \\
    \hline
    \rowcolor{highlight}
    ol\_blur - g\_transp & certain.aligned  & -0.941 & -1.504 & -0.381 \\
    \rowcolor{highlight}
    ol\_blur - g\_transp & certain.opposite &  1.197 &  0.521 &  1.846 \\
    \rowcolor{highlight}
    ol\_blur - g\_transp & danger.aligned   &  2.459 &  1.839 &  3.038 \\
    \rowcolor{highlight}
    ol\_blur - g\_transp & danger.opposite  & -2.496 & -3.045 & -1.922 \\
    \hline
    ol\_color - g\_transp & certain.aligned  &  0.432 & -0.020 &  0.890 \\
    ol\_color - g\_transp & certain.opposite & -0.363 & -1.005 &  0.272 \\
    \rowcolor{highlight}
    ol\_color - g\_transp & danger.aligned   &  3.184 &  2.600 &  3.731 \\
    \rowcolor{highlight}
    ol\_color - g\_transp & danger.opposite  & -3.207 & -3.751 & -2.613 \\
    \hline
    ol\_transp - g\_transp & certain.aligned  &  0.048 & -0.447 &  0.538 \\
    ol\_transp - g\_transp & certain.opposite & -0.248 & -0.888 &  0.396 \\
    ol\_transp - g\_transp & danger.aligned   &  0.388 & -0.221 &  1.001 \\
    \rowcolor{highlight}
    ol\_transp - g\_transp & danger.opposite  & -1.009 & -1.590 & -0.434 \\
    \hline
    \rowcolor{highlight}
    text - g\_transp & certain.aligned  &  0.539 &  0.099 &  0.986 \\
    text - g\_transp & certain.opposite & -0.324 & -0.960 &  0.315 \\
    \rowcolor{highlight}
    text - g\_transp & danger.aligned   &  2.005 &  1.353 &  2.633 \\
    \rowcolor{highlight}
    text - g\_transp & danger.opposite  & -1.872 & -2.463 & -1.260 \\
    \hline
    g\_blur - ol\_transp & certain.aligned  & -0.276 & -0.805 &  0.245 \\
    g\_blur - ol\_transp & certain.opposite &  0.292 & -0.348 &  0.929 \\
    g\_blur - ol\_transp & danger.aligned   &  0.014 & -0.628 &  0.649 \\
    g\_blur - ol\_transp & danger.opposite  &  0.330 & -0.304 &  0.961 \\
    \hline
    g\_color - ol\_transp & certain.aligned  &  0.147 & -0.321 &  0.626 \\
    g\_color - ol\_transp & certain.opposite &  0.494 & -0.171 &  1.146 \\
    \rowcolor{highlight}
    g\_color - ol\_transp & danger.aligned   &  3.297 &  2.748 &  3.812 \\
    \rowcolor{highlight}
    g\_color - ol\_transp & danger.opposite  & -2.207 & -2.834 & -1.534 \\
    \hline
    g\_transp - ol\_transp & certain.aligned  & -0.048 & -0.538 &  0.447 \\
    g\_transp - ol\_transp & certain.opposite &  0.248 & -0.396 &  0.888 \\
    g\_transp - ol\_transp & danger.aligned   & -0.388 & -1.001 &  0.221 \\
    \rowcolor{highlight}
    g\_transp - ol\_transp & danger.opposite  &  1.009 &  0.434 &  1.590 \\
    \hline
    \rowcolor{highlight}
    ol\_blur - ol\_transp & certain.aligned  & -0.989 & -1.547 & -0.424 \\
    \rowcolor{highlight}
    ol\_blur - ol\_transp & certain.opposite &  1.445 &  0.774 &  2.085 \\
    \rowcolor{highlight}
    ol\_blur - ol\_transp & danger.aligned   &  2.071 &  1.439 &  2.671 \\
    \rowcolor{highlight}
    ol\_blur - ol\_transp & danger.opposite  & -1.487 & -2.120 & -0.821 \\
    \hline
    ol\_color - ol\_transp & certain.aligned  &  0.384 & -0.070 &  0.847 \\
    ol\_color - ol\_transp & certain.opposite & -0.115 & -0.751 &  0.510 \\
    \rowcolor{highlight}
    ol\_color - ol\_transp & danger.aligned   &  2.796 &  2.187 &  3.368 \\
    \rowcolor{highlight}
    ol\_color - ol\_transp & danger.opposite  & -2.198 & -2.839 & -1.508 \\
    \hline
    \rowcolor{highlight}
    text - ol\_transp & certain.aligned  &  0.491 &  0.058 &  0.933 \\
    text - ol\_transp & certain.opposite & -0.076 & -0.705 &  0.551 \\
    \rowcolor{highlight}
    text - ol\_transp & danger.aligned   &  1.617 &  0.939 &  2.268 \\
    \rowcolor{highlight}
    text - ol\_transp & danger.opposite  & -0.863 & -1.544 & -0.172 \\
\label{tab:intuitest}
\end{longtable}
\end{center}

\begin{table}[H]
\caption{\rev{Preregistered hypotheses and a summary of their evidence.}}
\centering
\renewcommand{\arraystretch}{2.0} 
{\color{black}
\begin{tabular}{@{} p{1.2cm} p{8.5cm} N{3.5cm} @{}}
\toprule
 & Preregistered Hypotheses & Evidence \\
\midrule

\textit{H1} & Participants' likelihood of shutting off the electric grid would be more strongly affected by low qualitative confidence than by high qualitative confidence or no additional confidence information. & strong evidence to support \\[1.5em]

\textit{H2a} & Visualizations that express qualitative confidence using glyphs would exhibit a Qual. Confidence $\times$ \% Crossing interaction. In other words, low qualitative confidence will lead to a meaningfully different likelihood of turning off the circuit in comparison to high qualitative confidence, and that the \% Crossing condition would meaningfully change this difference. We did not hypothesize about the effects of medium confidence. & moderate evidence to support \\[1.5em]

\textit{H2b} & A three-way Qual. Confidence $\times$ \% Crossing $\times$ Confidence Encoding interaction. Specifically, that the Qual. Confidence $\times$ \% Crossing effect described above would vary as a function of forecasts' Confidence Encoding (blur, transparency, and color hue glyphs, and text). & no evidence to support \\[1.5em]

\textit{H3a} & Visualizations that encode confidence directly on top of forecast CIs (H3a) would exhibit a Qual. Confidence $\times$ \% Crossing interaction. In other words, we hypothesized we would see an effect in which low qualitative confidence led to a meaningfully different likelihood of turning off the circuit in comparison to high qualitative confidence, and that the \% Crossing condition would meaningfully change this difference. & moderate evidence to support \\[1.5em]

\textit{H3b} & A three-way Qual. Confidence $\times$ \% Crossing $\times$ Confidence Encoding interaction. Specifically, that the Qual. Confidence $\times$ \% Crossing effect described above would vary as a function of forecasts' Confidence Encoding (overlaid blur, transparency, and color hue encodings, and text). & no evidence to support \\

\bottomrule
\end{tabular}
}
\label{tab:hypotheses}
\end{table}

\begin{table}[H]
\caption{\rev{Summary of results by encoding, directional effects at \% crossing condition in comparison to text, and intuitiveness of uncertainty and danger. N indicates no meaningful difference between the encoding and text.}}
\centering
\small 
\renewcommand{\arraystretch}{1.4} 
{\color{black}
\begin{tabularx}{\textwidth}{
    |>{\raggedright\arraybackslash}p{2.2cm} 
    | >{\raggedright\arraybackslash}X 
    | >{\raggedright\arraybackslash}X 
    | >{\raggedright\arraybackslash}X 
    | >{\raggedright\arraybackslash}X 
    !{\vrule width 1.5pt} 
    >{\raggedright\arraybackslash}X 
    | >{\raggedright\arraybackslash}X |
}
\hline
\multicolumn{1}{|c|}{} & \multicolumn{4}{c!{\vrule width 1.5pt}}{Effect at \% crossing} & \multicolumn{2}{c|}{Intuitiveness} \\
\cline{2-5}\cline{6-7} 
\multicolumn{1}{|c|}{Encoding} & \multicolumn{1}{c|}{5\%} & \multicolumn{1}{c|}{25\%} & \multicolumn{1}{c|}{50\%} & \multicolumn{1}{c!{\vrule width 1.5pt}}{75\%} & \multicolumn{1}{c|}{Uncertainty} & \multicolumn{1}{c|}{Danger} \\
\hline

Transparency glyph &
Low confidence leads to slightly more cautious behaviour than high confidence, which is not the case with text &
Low confidence leads to more cautious behaviour than high confidence, which is not the case with text &
Low confidence does not lead to less cautious behaviour than high confidence, as it does with text &
Low confidence does not lead to less cautious behaviour than high confidence, as it does with text &
Highly logical to convey uncertainty, but less so than text &
Moderately illogical to convey danger, less logical than text and color hue \\
\hline

Blur glyph & N & N & N & N &
Highly logical to convey uncertainty, but less so than text &
Moderately illogical to convey danger, less logical than text and color hue \\
\hline

Color hue glyph & N & N & N & N & N &
Highly logical to convey danger, more so than all other encodings \\
\noalign{\hrule height 1.5pt} 

Blur overlaid &
Low confidence leads to slightly more cautious behaviour than high confidence, which is not the case with text &
Low confidence leads to more cautious behaviour than high confidence, which is not the case with text &
Low confidence does not lead to less cautious behaviour than high confidence, as it does with text &
Low confidence does not lead to less cautious behaviour than high confidence, as much as with text &
Slightly logical to convey uncertainty, but less so than all other encodings &
Moderately logical to convey danger, less logical than all other encodings \\
\hline

Transparency overlaid & N & N & N & N &
Highly logical to convey uncertainty, but less so than text &
Slightly illogical to convey danger, more so than transparency \\
\hline

Color hue overlaid & N & N & N & N & N &
Moderately logical to convey danger, more so than all other encodings \\
\hline

\end{tabularx}
}

\label{tab:encoding-summary}
\end{table}

\end{document}